\documentclass[preprint,pre,superscriptaddress,aps,nofootinbib,10pt]{revtex4-2}

\usepackage{amsmath}
\usepackage{graphicx}
\usepackage{dsfont}
\usepackage{amssymb}
\usepackage{bm}
\usepackage[hidelinks,colorlinks=true,linkcolor=blue,citecolor=blue,urlcolor=blue]{hyperref}
\usepackage{orcidlink}

\begin{document}
\title{Staggering domino-like blast front motion in a one-dimensional cold gas}

\author{Taras Holovatch\orcidlink{0009-0005-2953-8730}}
\email{holovatch@gmail.com}
\affiliation{Yukhnovskii Institute for Condensed Matter Physics, National Academy of Sciences of Ukraine, 79011 Lviv, Ukraine}
\affiliation{${\mathbb L}^4$ Collaboration \& Doctoral College for the Statistical Physics of Complex Systems,\\ Leipzig-Lorraine-Lviv-Coventry, Europe} 

\author{Yuri Kozitsky\orcidlink{0000-0002-4320-8835}}
\email{jurij.kozicki@mail.umcs.pl}
\affiliation{Institute of Mathematics, Maria Curie-Sk\l odowska University, {20-031} {Lublin}, {Poland}}

\author{Krzysztof Pilorz\orcidlink{0000-0003-2596-3260}}
\email{krzysztof.pilorz@mail.umcs.pl}
\affiliation{Faculty of Mathematics and Information Technology,
	Lublin University of Technology, {20-618} {Lublin}, {Poland}}

\author{Yurij Holovatch\orcidlink{0000-0002-1125-2532}}
\email{hol@icmp.lviv.ua}
\affiliation{Yukhnovskii Institute for Condensed Matter Physics, National Academy of Sciences of Ukraine, 79011 Lviv, Ukraine}
\affiliation{${\mathbb L}^4$ Collaboration \& Doctoral College for the Statistical Physics of Complex Systems,\\ Leipzig-Lorraine-Lviv-Coventry, Europe} 
\affiliation{Centre for Fluid and Complex Systems, Coventry University, Coventry, CV1 5FB, United Kingdom}
\affiliation{Complexity Science Hub, 1030 Vienna, Austria} 
\date{\today}

\begin{abstract} 
One-dimensional alternating particle systems are widely used to study interconnections between the hydrodynamics 
of blast waves in a gas-like medium and the Newtonian dynamics of its corpuscular 
constituents.  In this article, we study the model in which point particles with masses
$m,\mu, m, \mu, \dots$,  $(m \geq \mu)$ are distributed on the positive half-line $\mathds{R}_{+}$. Their dynamics are initiated by
giving a positive velocity to the leftmost particle; in its course, the particles undergo
elastic collisions. For this model with $m/\mu=2$, it has previously been established that the dynamics that start from random initial positions are consistent with predictions based on Euler's hydrodynamic equation. In particular, they have the following properties: (i) the position of the rightmost particle (shock
front) evolves as $t^\delta$ with $\delta <1$; (ii) recoiled particles behind the front enter the negative half-axis; (iii) particles with locations $x\leq 0$ move ballistically and eventually take
over the total energy of the system. 
In this paper, we present numerical and analytical results for the dynamics of this model with nonrandom (typically equidistant) initial positions and various values of $m/\mu$. For $m/\mu=2$ and equidistant initial positions, our results
qualitatively agree with those just mentioned. At the same time, we found an infinite family of numbers $\{\mathcal{M}_k\}$ such that, for $m/\mu = \mathcal{M}_k$, the hydrodynamic behavior mentioned changes drastically to the following. At each moment, only a single triplet $m, \mu, m$ is in motion, whereas all other particles are at rest. In the moving triplet, 
two heavy particles move to the right, while the lighter particle oscillates between them, transferring energy and momentum to the right. As a result, the shock front moves ballistically with an average velocity equal to the initial one. Such a `staggering domino-like' picture is obtained as an exact solution, which yields, in particular, explicit formulas for $\mathcal{M}_k$ and the particle velocities and positions. Its details and importance in a wider context are also discussed.     
\end{abstract}
\maketitle 

\section{Introduction} \label{I}

The relationship between the macroscopic dynamics of matter, viewed as a continuous medium, 
and that of its microscopic corpuscular constituents is a paramount challenge in physics \cite{Spohn80,Lebowitz21}. Its prominent 20th-century formulation is Hilbert's sixth problem, which specifically calls 
for `developing mathematically the limiting processes... which lead from the atomistic view to 
the laws of motion of continua' \cite{Gorban18}. This applies to the transport of energy
in a cold gas following an explosion -- an instant injection of
energy at a certain point.
In this paper, we show that a typical and relatively simple 
model of this phenomenon may exhibit surprisingly counterintuitive behavior. 

The model we consider 
is an infinite collection of point particles located in the positive part of the one-dimensional space $\mathds{R}$, the dynamics of which is initiated by giving 
the leftmost particle a unit positive velocity, whereas all other particles remain motionless; see, e.g., 
\cite{Antal08,Chakraborti22}. This motion is then transferred to the right through elastic collisions, resulting in a blast wave moving in the positive direction. Recoiled particles behind the shock front can form a splatter that moves to the left.  
Let $\mathcal{R}(t)$ denote the location of the rightmost moving particle, the shock front. If the particles have equal masses, the dynamics is of the domino type, and the front moves with unit velocity, i.e., $\mathcal{R}(t) = t$, which is contrary to predictions based on hydrodynamic equations. Usually, this kind of degeneracy is eliminated by varying the masses of particles. In the simplest case, 
two different values, $m$ and $\mu\leq m$, are periodically alternated. In \cite{Chakraborti22}, such a model was studied with $m/\mu = 2$ and random initial positions of the particles. It was found that: (i) $\mathcal{R}(t) \sim t^\delta$ with $\delta<1$, i.e., the shock front propagates in a hydrodynamic way, showing both qualitative agreement and quantitative consistency with the exponent predicted by the Euler equation; (ii) recoiled particles behind the front (splatter) enter
the negative half-line and
eventually take over the whole energy of the system. 
\newpage

\begin{figure}[h]
	\centerline{\includegraphics[width=0.49\linewidth]{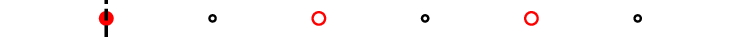}}
    
    \centerline{\textbf{(a)}}
	\centerline{\includegraphics[width=0.49\linewidth]{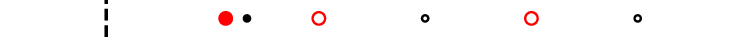}}
    \centerline{\textbf{(b)}}
    \centerline{\includegraphics[width=0.49\linewidth]{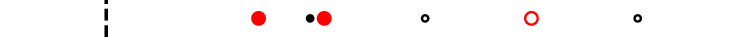}}
    \centerline{\textbf{(c)}}
	\centerline{\includegraphics[width=0.49\linewidth]{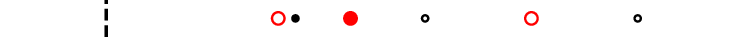}}
    \centerline{\textbf{(d)}}
	\centerline{\includegraphics[width=0.49\linewidth]{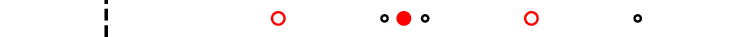}}
    \centerline{\textbf{(e)}}    
    \caption{Snapshots of the particle motion for \(m=\mathcal{M}_2=2+\sqrt{5}\) and an equidistant initial arrangement, with time increasing from top to bottom. Particles are treated as point-like; their shapes in the figure are for visualization purposes only. Large red and small black circles represent heavy (\(m=\mathcal{M}_2\)) and light (\(\mu=1\)) particles, respectively, with filled circles indicating those currently in motion. \textbf{(a)} At \(t=0\), all particles are equidistant, and the leftmost particle 0 moves to the right. \textbf{(b, c)} Snapshots following the \(0\leftrightarrow 1\) collision, and before and after the \(1\leftrightarrow 2\) collision, respectively. \textbf{(d)} The state after the second \(0\leftrightarrow 1\) collision, where particle 1 moves to the right and particle 0 has come to rest. \textbf{(e)} Completion of the first cycle: particle 2 moves toward particle 3 to initiate the dynamics within the next triplet, \(2,3,4\). We refer to this kind of dynamics as a staggering domino-like motion.}

	\label{fig:triplets}
\end{figure}

In this paper, we study the dynamics of this model for various values of $m/\mu$ and for nonrandom initial particle positions, with the leftmost particle having mass $m$. We found a countably infinite family $\{\mathcal{M}_k: k \geq 1\}$, $\mathcal{M}_{k+1}>\mathcal{M}_k > \mathcal{M}_1 = 1$, $\mathcal{M}_k \to +\infty$, such that for $m/\mu = \mathcal{M}_k$ and the initial particle positions satisfying a condition dependent on $k$, the dynamics of the system amounts to the motion of triplets $m$, $\mu$, $m$, in the course of which all other particles are motionless. In this motion, two heavy particles in the triplet move rightward, whereas the lighter particle oscillates between them, thereby transferring energy and momentum to the right. An equidistant arrangement of the initial positions satisfies the aforementioned condition for all $k$.
The motion starts from the leftmost particle and its two consecutive neighbors; see Fig. \ref{fig:triplets}. This triplet, say particles $0,1,2$,  undergoes $k$ rounds of mutual collisions $0\leftrightarrow  1$ and $1 \leftrightarrow 2$,
in which particles $0$ and $2$ collide $k$ times, while particle $1$ collides $2k$ times. At the end of the $k$-th round, particles $0$ and $1$ stop, and particle $2$ moves rightward with unit velocity and thus initiates the motion 
of the triplet $2,3,4$, which repeats the same $k$ rounds of mutual collisions, and so on. As a result, the shock front moves at an average velocity of $1$, and the splatter is absent. We refer to this kind of dynamics as a staggering domino-like motion, which we describe in formulas and numerically. In particular, we present explicit expressions for $\mathcal{M}_k$ and the particle velocities and positions. 

The remainder of the paper is organized as follows. In the next section \ref{II}, we provide a brief review of the literature connecting the Newtonian dynamics of interacting particles to the hydrodynamics of wave propagation in continuous media.  Section \ref{III} defines the model and introduces the observables which are the objects of our analysis. The main results are presented in Sections \ref{IV} and \ref{V}. In Section \ref{IV}, we describe and discuss the evolution of the relevant observables obtained by the molecular dynamics simulations that we performed. Then we outline the derivation of explicit formulas for the particle velocities and positions, Section \ref{V}. We conclude the presentation with a summary and an outlook, Section \ref{VI}. A detailed derivation of the formulas is deferred to the Appendices. Some of our results were announced in a Letter \cite{Holovatch25b}.

\section{Blast and splash in a cold gas} \label{II}

The study of blast waves in continuous media has a relatively short history, emerging from the need to predict hydrodynamic effects following a massive, instantaneous energy release in a small volume \cite{Taylor50a,Sakurai65}. The arguments coming from the famous `Taylor-von Neumann-Sedov solution' of the Euler equation \cite{Taylor50a,Sedov59,Neumann63}, predict a power-law large-time asymptotic of the shock front propagation 
\begin{equation}\label{1.1}
\mathcal{R}(t) \sim t^\delta\, ,
\end{equation}
with an exponent dependent on the spatial dimension
$d$ \cite{Zeldovich02,Landau87}:
\begin{equation}\label{1.2}
\delta = 2/(d+2)\, .
\end{equation}
Other observables characterizing the blast wave motion also exhibit power-law asymptotics, as discussed below. 
This problem was originally motivated by the need to understand the mechanical effects of a large-scale nuclear explosion \cite{Taylor50a,Taylor50b}. Similar phenomena are now observed in high-energy laser pulses in gas jets \cite{Edwards01}, plasma \cite{Edens04}, various gases \cite{Moore05}, and astrophysical events such as supernova explosions \cite{Ostriker88,Tang17}.

Numerous studies \cite{Du95,Antal08,Jabeen10,Barbier15,Joy21a,Joy21b,Kumar25a,Kumar25c}, both analytical and numerical, have addressed how the hydrodynamic propagation of blast waves in a continuous medium is shaped by the Newtonian dynamics of its constituent particles. Specifically, these works examine systems of initially stationary, elastically interacting particles, focusing on the dynamics triggered by imparting an initial velocity to a single particle. In this formulation, the problem can be viewed as an example of $N$-particle billiards \cite{Sinai70,MacKay87,Tabachnikov05}. However, unlike standard billiard problems involving a few particles in a confined space, this model deals with the multi-particle dynamics of so-called `statistical billiards' \cite{Bunimovich00}. Further extensions of this model incorporate inelastic collisions and dissipative media \cite{Du95,Jabeen10,Barbier15}.

A special case of this problem is the propagation of blast waves in one dimension ($d=1$), which is the focus of the present work. This choice is motivated not only by the fact that one-dimensional systems are more amenable to analytical study and large-scale simulation than their more realistic two- or three-dimensional counterparts, but also by unique physical features inherent to the phenomenon itself. Indeed, relation (\ref{1.1}) implies that the time dependence of the observables characterizing blast wave propagation can be expressed in terms of the wavefront radius $\mathcal{R}$. Consequently, the corresponding scaling relations are expected to hold across different dimensions $d$ and can be rigorously verified in the simpler $d=1$ case. Furthermore, the applicability of one-dimensional models is particularly pronounced across diverse physical contexts. In our case, the particle model can be mapped onto spin or interface models, establishing a connection to the Kardar–Parisi–Zhang (KPZ) universality class \cite{Grassberger02,Hurtado16,Lepri20,Grassberger23}. However, a defining constraint of such one-dimensional systems is the invariance of the particle sequence during their time evolution \cite{Hurtado06}.

As already mentioned in Section \ref{I}, for a one-dimensional chain of elastically interacting particles of equal mass, the dynamics initiated by setting a single particle in motion is trivial and domino-like: with each collision, energy and momentum are transferred entirely to the neighboring particle, ensuring that at any given time, only one particle is in motion.
Therefore, to study blast wave propagation in a one-dimensional chain, an alternating hard particle (AHP) model is typically employed \cite{Antal08,Hurtado06,Ganapa21,Chakraborti21a}. 
Similar models are used to analyze heat transport and hydrodynamics in one dimension, see, e.g., \cite{Dhar01,Garrido01,Grassberger02,Casati03,Cipriani05,Hurtado16,Lepri20} and references therein.
 In such a model, one considers a chain of point particles with two different masses that alternate periodically. 
As shown in Refs. \cite{Ganapa21,Chakraborti21a}, an instantaneous localized release of energy in the bulk of an AHP gas evolving according to Newtonian dynamics leads to the development of a blast wave. Its evolution is in remarkable agreement with Euler hydrodynamics, despite some deviations in a small core region \cite{Kumar25a}.

A symmetric blast in the bulk of a gas-like medium typically results in a symmetric shock wave propagating in 
both directions. This symmetry is broken when the medium occupies a half-space and the blast occurs 
at its boundary \cite{Antal08,Chakraborti22}. In this configuration, the shock wave penetrates the gas as it would in a bulk system, though back-scattered particles eventually form what is termed a \emph{splatter}.
The long-time behavior of an AHP gas model of this kind exhibits remarkable features \cite{Chakraborti22}: (i) the shock front motion is governed by Euler's hydrodynamics, and (ii) the splatter moves backward in a ballistic way. In this asymmetric case, the shock front scaling exponent $\delta$ (see (\ref{1.1})) differs from its value $\delta=2/3$ in the one-dimensional symmetric case, cf. Eq. (\ref{1.2}). 
Its value, given in Table \ref{tab1}, was calculated in \cite{Chakraborti22} using scaling arguments applied to the corresponding hydrodynamic equation and then confirmed by molecular (Newtonian) dynamics calculations with \textit{random} initial positions of the particles. 
The exponents corresponding to other observables, obtained in a similar way, are also listed in Table \ref{tab1}. Their physical significance is discussed below. In this work, we focus on the Newtonian dynamics results -- both analytic and numeric -- corresponding to \textit{deterministic} initial configurations.

\begin{table}[t]
	\caption{Exponents governing the large-time asymptotics for the one-dimensional blast and splash problem obtained in
	\cite{Chakraborti22}. They correspond to the shock front \eqref{1.1}, $\delta$, 
    the total energy, $\beta$ and momentum, $\gamma$ in the blast and splash regions, 
    and the total number of collisions, $\eta$ \eqref{indices}.
	}\label{tab1}
	\begin{center}
		\tabcolsep1.2mm
		\begin{tabular}{|c|c|c|c|}
			\hline
			 $\delta$ & $\beta$ & $\eta$ & $\gamma$  \\			
			\hline
			 0.6279520544 & 0.11614383675 & 1.255904109 & 0.2559041088 \\
			\hline 
		\end{tabular}
	\end{center}
\end{table}

\section{Setup and Key Observables} \label{III} 

We study the dynamics of a system of point particles placed in the positive part of $\mathds{R}$. 
Throughout, by $\mathds{R}$ and $\mathds{N}$, we denote the sets of real and natural numbers, respectively. We also use the set $\mathds{N}_0 = \mathds{N} \cup \{0\}$, consisting of all nonnegative integers. At $t=0$, the leftmost particle starts moving rightward with a unit velocity, while all other particles
remain motionless; see Fig. \ref{fig:triplets}\textbf{a}.
The only interparticle interaction is an elastic collision in which the colliding particles preserve total momentum and kinetic energy. Therefore, if particles $a$ and $b$ (having masses $m_a$, $m_b$ and velocities $u_a$, $u_b$, respectively) collide, then their velocities $v_a$, $v_b$ after collision are given by the formula
\begin{equation}
 \label{1a}
 v_a= \frac{m_a- m_b}{m_a+m_b} u_a + \left(1- \frac{m_a- m_b}{m_a+m_b} \right) u_b,
\end{equation}
and its version for $v_b$, which one obtains by interchanging the indices in \eqref{1a}. 

For convenience, we number the particles by nonnegative integers $l\in \mathds{N}_0$. 
The dynamics will start at the initial positions of the particles $x_l (0)\in \mathds{R}_{+}:= [0,+\infty)$. As a prototype example, the equidistant arrangement  $x_l(0) = l$ can be kept in mind.
The particle masses alternate between two values $1$ and $m\geq 1$. The leftmost particle has mass $m$; that is, $m_{2l} = m$ and $m_{2l+1} =1$ for all $l\in \mathds{N}_0$. Without loss of generality, we set $u_0(0)=1$. The particles gradually collide and alternate their positions $x_l(t)$, and velocities $u_l(t)$. In particular, some of them can enter the negative half-line $(-\infty, 0)$. Although the whole system of particles is infinite, at each $t>0$, only finitely many particles have nonzero velocities. However, the number of such particles may grow with time. For a given $t>0$, let $l_t\in \mathds{N}_0$ be such that $u_{l_t}(t)>0$, while $u_l(t)=0$ for all $l>l_t$. Then
\begin{equation}
 \label{1R}
 \mathcal{R}(t) = x_{l_t} (t)
\end{equation}
is the position,  at time $t$, of the rightmost moving particle, the shock front. Together with the observable in \eqref{1R}, we shall study the total number of collisions up to time $t$, denoted $\mathcal{C}(t)$, the total energy at time $t$ of the particles located in $\mathds{R}_{+}$, denoted $\mathcal{E}_{x\geq 0} (t)$, and the total momentum of the particles located in $(-\infty, 0)$ at time $t$, denoted $\mathcal{P}_{x< 0}(t)$. That is,
\begin{equation}
\label{1E}
\mathcal{E}_{x\geq 0} (t) = \sum_{l: x_l(t)\geq 0} m_l u_l^2(t) /2, \quad \mathcal{P}_{x<0} (t) = \sum_{l: x_l (t) < 0} m_l u_l (t).
\end{equation}
For $m=1$, the shock front motion becomes degenerate for an arbitrary initial arrangement. It is of domino type since $\mathcal{R}(t)=t$, $\mathcal{E}_{x\geq 0} (t) = 1/2$, $\mathcal{P}_{x<0} (t)=0$, and $\mathcal{C}(t) = [t]$, the integer part of $t>0$. In this case, there is only one moving particle for each $t>0$. For $m>1$, the character of the dynamics becomes different. In particular, the initial energy $m/2$ is now distributed among numerous moving particles. As a measure of the deviation of such dynamics from the degenerate domino-like motion, we use the Shannon entropy 
\begin{equation}
\label{1H}
\mathcal{H}(t) = - \sum_{l\geq 0}p_l (t) \log p_l (t), \qquad p_l (t) = u_l^2 (t) m_l/m,
\end{equation}
with the conventions $\log 2 =1$ and $0 \log 0 =0$. In fact, $p_l(t)$ is the normalized energy of particle $l$ at time $t$. We take these energies normalized as in \eqref{1H} to secure equality $\sum_{l} p_l (t) =1$, which is standard in this case. Note that $\mathcal{H}(t) =0$ for all $t>0$  for the degenerate domino dynamics. Hence, the higher the value of $\mathcal{H}(t)$, the more uniform the distribution of the initial energy between the particles.

\section{Molecular Dynamics Simulations}\label{IV}

Here, we present and discuss the numerical calculation of the aforementioned observables based on the rule in \eqref{1a} and performed by molecular dynamics (MD) simulations. In the course of such simulations, we study the dynamics in a window encompassing the first $N$ particles, with values of $N$ ranging in the interval $10^4-10^5$. In particular, this means that the observed dynamics are terminated when the shock front reaches particle $N-1$.
We denote the values of $\mathcal{E}_{x\geq 0} (t)$, $\mathcal{H}(t)$, $-\mathcal{P}_{x<0}(t)$, and $\mathcal{C}(t)$ at this moment as $\mathcal{E}_{\rm fin}(m)$, $\mathcal{H}_{\rm fin}(m)$, $\mathcal{P}_{\rm fin}(m)$, and $\mathcal{C}_{\rm fin}(m)$, explicitly indicating their dependence on $m$.
These values are calculated using event-driven Molecular Dynamics (MD) simulations. A heap data structure -- see \cite{Williams1964} -- is populated with all potential collisions at the moment of initialization; the simulation then proceeds by identifying and processing the earliest valid collision.

We begin by presenting the time evolution of the observables introduced above for 
$m = 2, 3,$ and 10. The dynamics of this model for $m = 2$ -- with initial positions sampled from a uniform ensemble -- were previously studied in \cite{Chakraborti22}, where it was shown that these observables exhibit the following large-time asymptotics:

\begin{equation}
 \label{indices}
 \mathcal{R}(t) \sim t^\delta, \quad \mathcal{E} _{x\geq 0}(t) \sim t^{-\beta}, \quad
 - \mathcal{P}_{x<0} (t) \sim t^\gamma, \quad
 \mathcal{C}(t) \sim t^\eta\, 
\end{equation}
with the exponents specified in Table \ref{tab1}. 
A characteristic feature of these dynamics is that, while all the energy of the system eventually transfers to the
region $x<0$, the momentum in both $x<0$ and $x\geq 0$ regions increases with the same asymptotics: 
$ \mathcal{P}_{x\geq 0} (t) \sim -  \mathcal{P}_{x<0} (t) \sim t^\gamma$, which is consistent with the law of momentum conservation $\mathcal{P}_{x\geq 0} (t)  +  \mathcal{P}_{x<0} (t) = m$.
Moreover, the scaling laws derived in \cite{Chakraborti22} yield the following relations between the exponents \eqref{indices} 
\begin{equation}
 \label{scaling}
 \eta = 2 \delta , \quad \delta= (2-\beta)/3\, , \quad \gamma=(1-2\beta)/3\, . 
\end{equation}

\begin{figure}[h]
	\centerline{ 
        \includegraphics[angle=0,origin=c,width=0.48\linewidth]{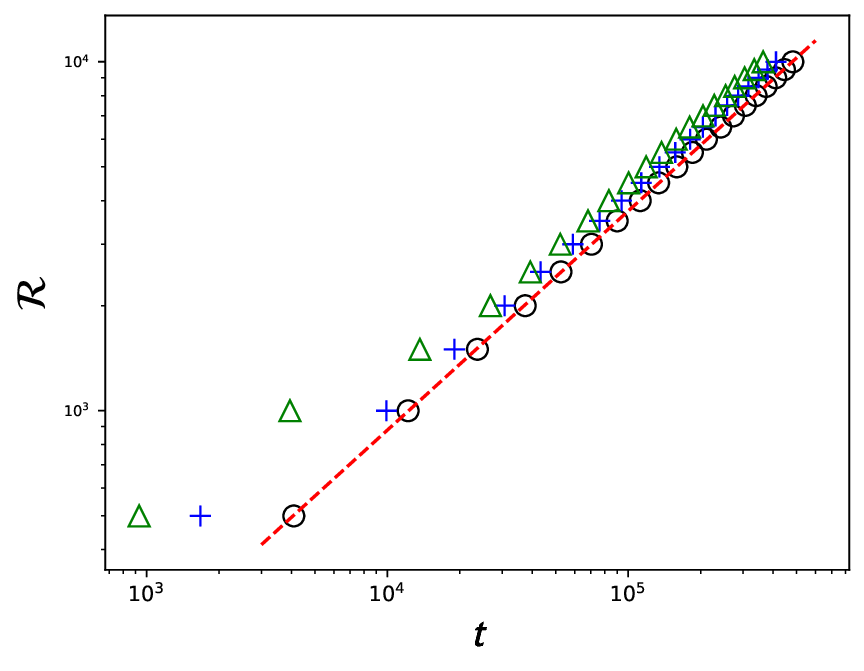}
		\includegraphics[angle=0,origin=c,width=0.52\linewidth]{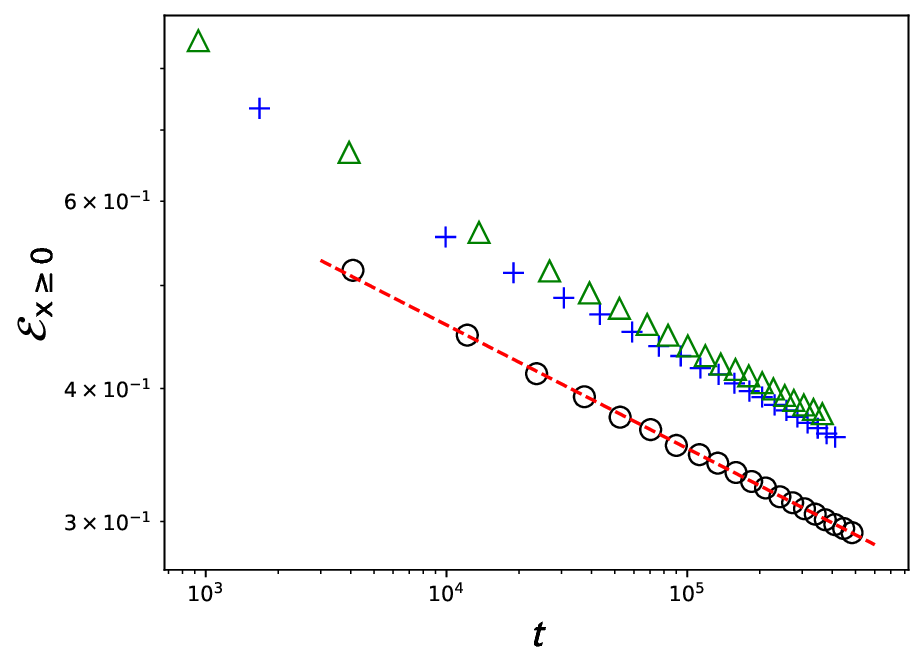}
	}
	\centerline{\textbf{(a)} \hspace{25em} \textbf{(b)}}
	\centerline{	
        \includegraphics[angle=0,origin=c,width=0.5\linewidth]{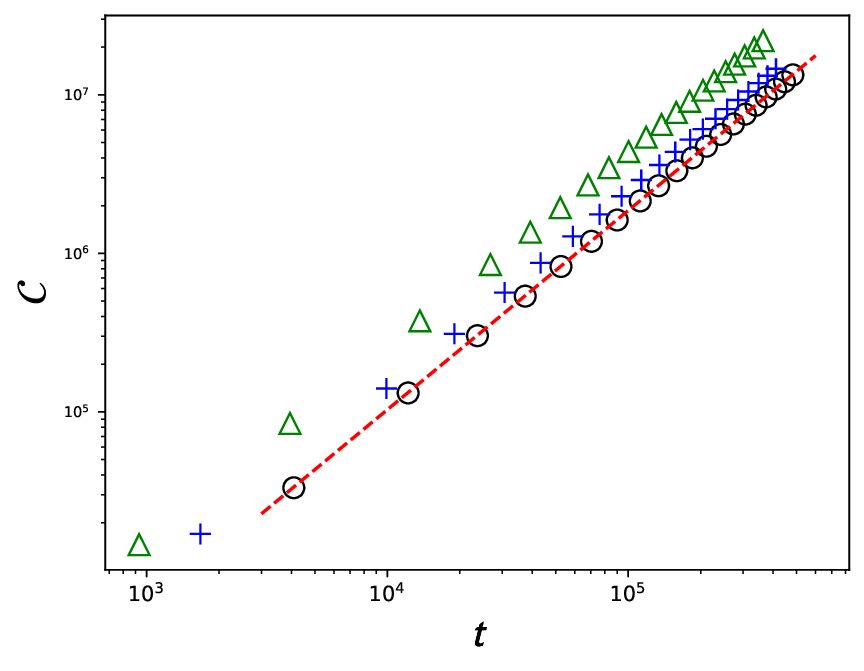}	\includegraphics[angle=0,origin=c,width=0.5\linewidth]{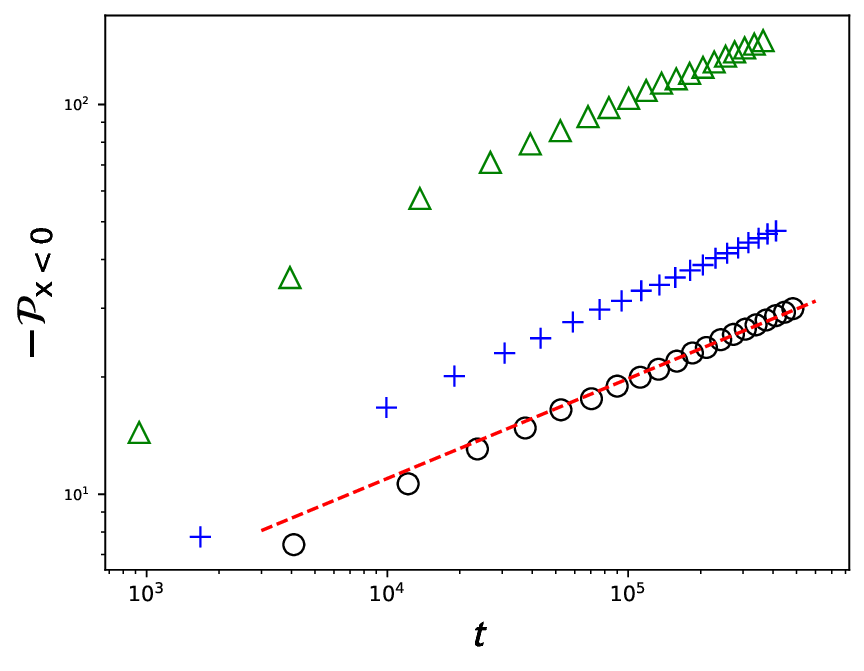}
	}
	\centerline{\textbf{(c)} \hspace{25em} \textbf{(d)}}
	\caption{Time dependence of: \textbf{(a)} blast front position ${\cal R}(t)$; \textbf{(b)} total energy of the particles in the $x \geq 0$ region ${\cal E}_{\rm{x} \geq 0}(t)$; \textbf{(c)} number of collisions ${\cal C}(t)$; \textbf{(d)} total momentum of the  particles in the $x < 0$ region ${\cal P}_{\rm{x}<0}(t)$. 
    Data points indicate the instances when the blast wave passes every
    $500\times k$-th particle, $k=1,\dots ,20$. 
    Black circles, blue pluses, and green triangles correspond to
$m=2$, $m=3$, and $m=10$, respectively. Red dashed lines show 
power-law fits with the exponents given in Table \ref{tab1}.}	
	\label{fig:powers}
\end{figure}

Our results for $m=2, 3, 10$ with equidistant initial positions $x_l(0)=l$ are shown in Fig. \ref{fig:powers}. These results demonstrate the emergence of power-law asymptotics with exponents consistent with the values listed in Table \ref{tab1}. The accuracy of these calculations depends on both the system size and the specific time window selected for the fit. For a more detailed discussion on universal and effective scaling exponents, we refer the reader to \cite{Holovatch25a}.

\begin{figure}[h]
	\centering
	\centerline{
        \includegraphics[angle=0,origin=c,width=0.5\linewidth]{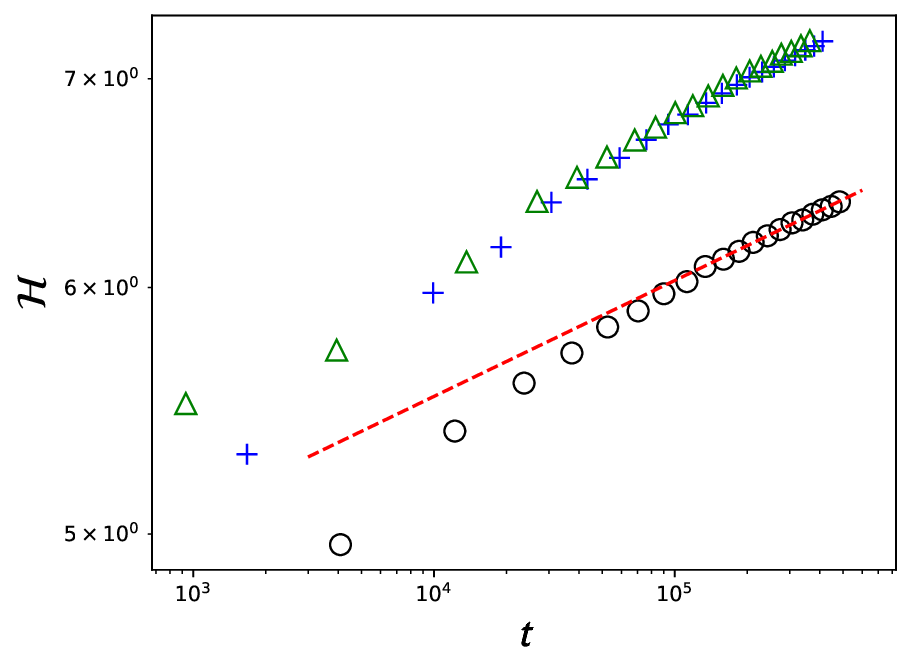}
	}
	\vspace{-3ex}
	\caption{Time dependence of the Shannon entropy ${\cal H}(t)$;
    symbols as in Fig. \ref{fig:powers}.
    Red dashed line shows a power-law fit with the exponent $0.037$.}
	\label{fig:shannon_power}
\end{figure}

In addition to the observables presented in  \eqref{indices}, we study the evolution of Shannon's entropy \eqref{1H}. According to the results presented in Fig. \ref{fig:shannon_power}, 
for the indicated values of $m$, it increases as
\begin{equation}
 \label{scaling_entropy}
 \mathcal{H}(t) \sim t^\alpha \, .
\end{equation}
Let us analyze this fact. According to the aforementioned momentum conservation law and \eqref{indices}, it follows that  
$\mathcal{P}_{ x\geq 0} (t) = m - \mathcal{P}_{ x< 0} (t) \sim t^\gamma$, which means that the quantity
\begin{equation}
\label{QQ}
\mathcal{Q}(t) := \sum_{l: x_l(t) \geq 0} m_l |u_l(t)| \geq \mathcal{P}_{ x\geq 0} (t) \sim t^\gamma
\end{equation}
also increases in time. At the same time, by \eqref{1E} and \eqref{indices} we have 
\[
\sum_{l: x_l(t) \geq 0} m_l u^2_l(t) \sim t^{-\beta}.
\]
Since the coefficients, here and \eqref{QQ}, take just two values $m_l=1,m$, the latter and \eqref{QQ} yield bounds for the convergence $u_l (t) \to 0$ as $t\to \infty$. To obtain similar bounds for all $u_l(t)$, we use the following arguments. By the energy conservation law, we have $\sum_{l\geq 0} m_l u_l^2(t) = m$. At the same time, by     
\eqref{scaling_entropy} and \eqref{1H}
\[
\mathcal{H} (t) = - \frac{2}{m}\sum_{l\geq 0} m_l u_l^2(t) \log |u_l (t)| + \frac{2}{m}\sum_{l: {\rm odd}} m_l u_l^2(t) \sim t^\alpha.
\]
By combining the latter with \eqref{QQ} we obtain
\begin{gather}
\label{QQ1}
\sum_{l\geq 0} m_l u_l^2 (t) = m, \qquad \sum_{l\geq 0} m_l |u_l (t)| \geq \mathcal{P}_{ x\geq 0} (t) \sim t^\gamma, \\ \nonumber \sum_{l\geq 0} m_l u^2_l (t) \log  |u_l (t)| \sim t^\alpha.
\end{gather}
Note that for $m=2$, $\alpha$ is much smaller than $\gamma$, which means that the increase of the latter sum is much slower than that of $\sum_{l\geq 0} m_l |u_l (t)|$.

As mentioned above, our main aim is to study the peculiarities of the dynamics of the model for different values of $m$. The authors of 
Ref. \cite{Chakraborti22} state that their results for $m=10$
and random initial positions are similar to those for $m=2$. 
Our simulations for equidistant initial positions with \(m=3\) and \(m=10\) confirm this behavior; see the data points represented by triangles and pluses in Fig. \ref{fig:powers}.

\begin{figure}[h]
	\centerline{ 
        \includegraphics[angle=0,origin=c,width=0.51\linewidth]{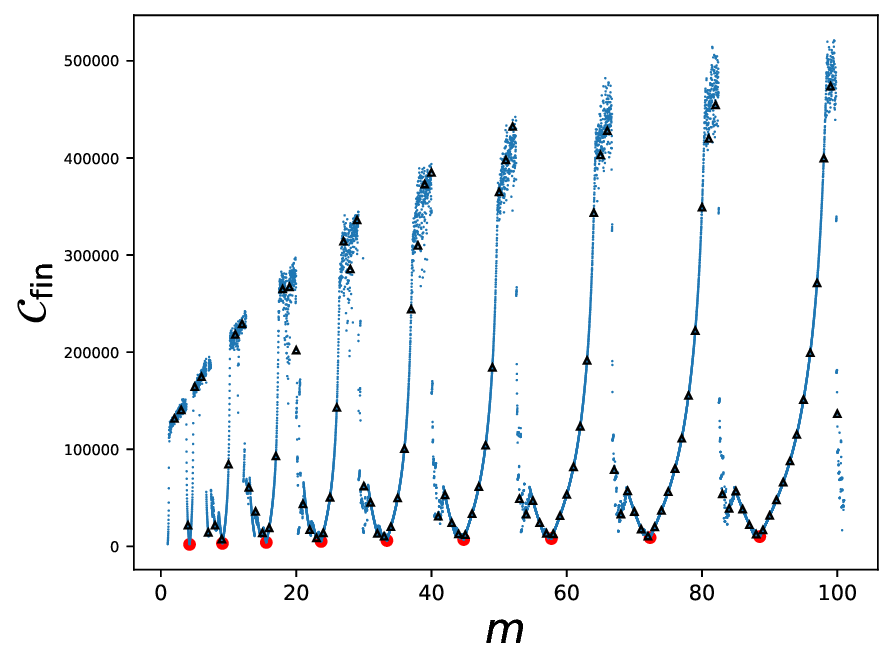}
		\includegraphics[angle=0,origin=c,width=0.49\linewidth]{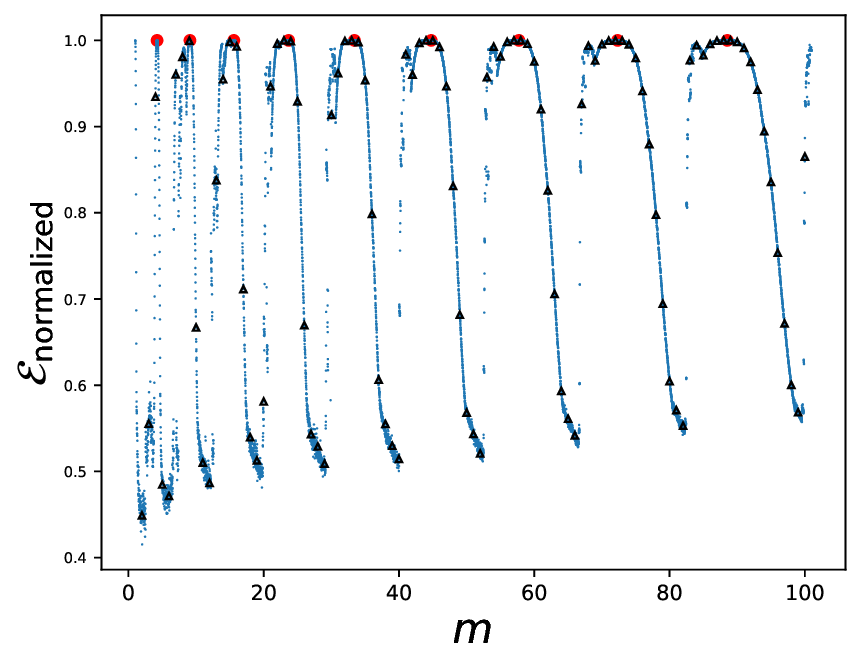}
	}
	\centerline{\textbf{(a)} \hspace{25em} \textbf{(b)}}
	\centerline{	
        \includegraphics[angle=0,origin=c,width=0.5\linewidth]{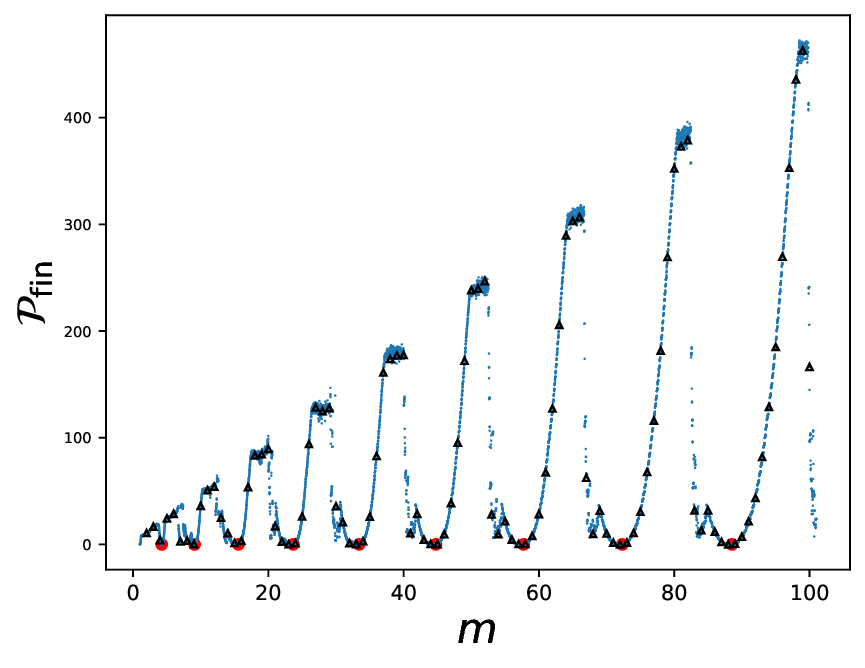}	\includegraphics[angle=0,origin=c,width=0.49\linewidth]{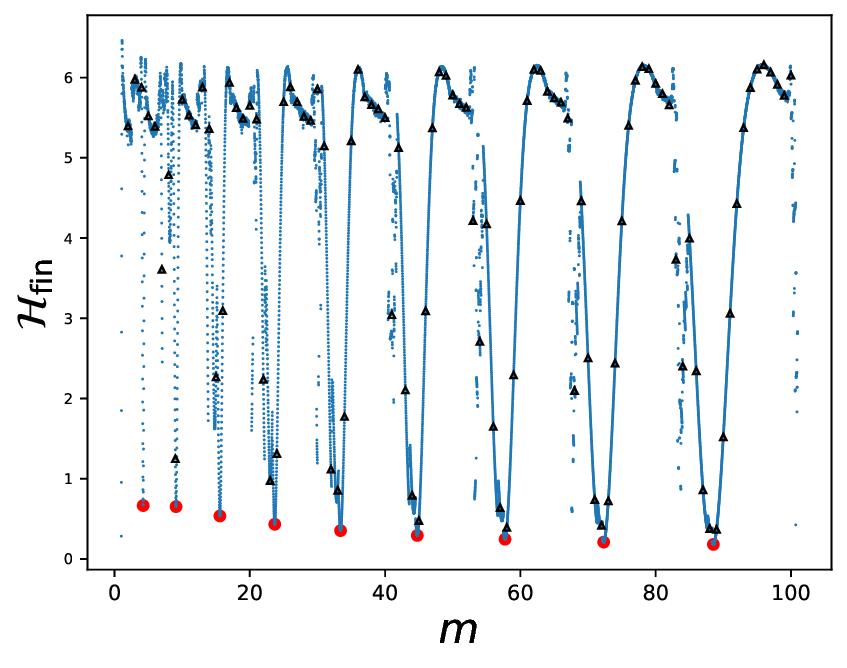}
	}
	\centerline{\textbf{(c)} \hspace{25em} \textbf{(d)}}
	\caption{Mass $m$ dependence of the observables at the moment the shock front reaches particle $l=N-1$, $N=10^4$:  \textbf{(a)} total number of collisions $\mathcal{C}_{\rm fin}$; \textbf{(b)} normalized energy $\mathcal{E}_{\rm normalized}=2\mathcal{E}_{\rm fin}/m$; \textbf{(c)} momentum $\mathcal{P}_{\rm fin}$; \textbf{(d)} Shannon entropy $\mathcal{H}_{\rm fin}$. Scattered blue dots: resolution $\Delta m=0.01$; black triangles: integer values of $m$; red disks: $m=\mathcal{M}_k$.}
    \label{fig:ratios}
\end{figure}

However, examining the behavior of these observables over a broader range of 
$m$ yields the following results (see Fig. \ref{fig:ratios}). 
For these specific simulations, the process is terminated once the shock 
front reaches $l=N-1$, with $N=10^4$. The significant feature of 
$\mathcal{C}_{\rm fin}(m)$, see panel (a), is that it periodically reaches minima at certain points, $m=\mathcal{M}_k$, with the amplitude of
oscillations increasing with $m$. At the very same values of $m$, the energy $\mathcal{E}_{\rm fin}(m)$ reaches its maximal value equal to the initial energy, $\mathcal{P}_{\rm fin}(m)$ hits zero, and the Shannon entropy gets close to zero as well; see panels (b), (c) and (d) of Fig. \ref{fig:ratios}. Note that the maximum values of the entropy seem to be the same for all observed values of $m$. These observations clearly suggest that for $m=\mathcal{M}_k$ and an equidistant initial arrangement the splatter never appears, and the shock front moves ballistically. Thus, the staggering domino-like dynamics mentioned above may be possible. In the following, we show in formulas that this is indeed the case, including the calculation of all $\mathcal{M}_k$; see \eqref{M33}. In this aspect, the object with which we deal can be considered as an exactly solvable model; cf. \cite{Baxter07}.  
In the next section, we outline our calculations, whereas their complete (rather lengthy) presentation is made in the Appendices.

\section{Staggering domino dynamics in formulas}\label{V}

\subsection{Beginning}
\label{VB}
Recall that we consider an infinite collection of point particles numbered $0,1,2, \dots$, initially located at points $\bar{x}_l$, $l \in \mathds{N}_0$, $\bar{x}_{l} < \bar{x}_{l+1}$ for all $l$, and $\bar{x}_l \to +\infty$. A prototype example is $\bar{x}_{l} =l$, where the initial positions are equidistant. The corresponding dynamics are characterized in Figs. \ref{fig:powers}, \ref{fig:shannon_power}, and \ref{fig:ratios}. Even-numbered particles are of mass $m\geq 1$, while odd-numbered particles are of unit mass. Together with $m$, we also use the parameter, cf. \eqref{1a},
\begin{equation}
\label{1}
\theta = \frac{m-1}{m+1},
\end{equation}
that takes values in $[0,1)$. First, we consider the triplet of particles $0,1,2$ alone. According to the conservation laws, the corresponding velocities satisfy
\begin{gather}
\label{M1a}
u_0 (t)+ u_2 (t) + u_1(t)/m = 1, \\[.2cm] \nonumber [u_0 (t)]^2 + [u_2 (t)]^2 + [u_1 (t)]^2/m = 1, 
\end{gather}
from which we readily deduce
\begin{equation}
\label{2}
u_0 (t) [1-u_0 (t)] + u_2 (t)[ 1-u_2(t)] = \frac{1}{m} u_1 (t)[u_1(t) -1], 
\end{equation}
and also $u_0(t)\leq 1$, $u_2 (t)\leq 1$. In the dynamics of the triplet $0,1,2$, particle 2 can only increase its velocity, hence $u_2(t)\geq 0$. Recall that $u_0(0) =1$ and $u_l (0)=0$ for $l=1,2$, and $x_3(0)=\bar{x}_3$. By \eqref{2}, if $u_0 (t) >0$, then $u_1(t)$ is negative or greater than 1. Consequently, the next collision is either $0\leftrightarrow 1$ or $1\leftrightarrow 2$. These two collisions will be repeated, one by one in that order, until $u_0(t)$ is strictly positive. The series of two such collisions -- first $0\leftrightarrow 1$, then $1\leftrightarrow 2$ -- will be called a \emph{round}. If $u_0 (t)>0$ after $0\leftrightarrow 1$, then $1\leftrightarrow 2$ occurs, provided at the moment of this collision, $x_2(t)< \bar{x}_3$. Otherwise, $2\leftrightarrow 3$ occurs before it. If $u_0 (t)$ becomes negative after $0\leftrightarrow 1$, then $1\leftrightarrow 2$ may not occur. Of course, all these scenarios are $m$-dependent. Keeping this in mind, we aim to find $\mathcal{M}_k$, $k\geq 2$, such that: 
\begin{itemize}
\item[($\mathcal{A}$)] The \underline{equidistant} initial arrangement $x_l (0) = l$, $l\in \mathds{N}_0$: for each $m=\mathcal{M}_k$, 
particles $2l,2l+1,2l+2$, $l\in \mathds{N}_0$, perform $k$ rounds of mutual collisions. After the last collision $2l+1\leftrightarrow 2l+ 2$, the following holds: 
\begin{itemize}
\item{($\mathcal{A}$1)} $u_{2l}(t) = u_{2l+1}(t) =0$, $u_{2l+2} (t) =1$; 
\item{($\mathcal{A}$2)} $x_{2l+2} (t)< x_{2l+3} (0) = 2l+3$ at the moment of this collision. All other particles are at rest.
\end{itemize}
\item[($\mathcal{B}$)]  The \underline{general} initial arrangement $x_l (0) = \bar{x}_l$, $l\in \mathds{N}$: for each $m=\mathcal{M}_k$, the particles perform the dynamics described in item ($\mathcal {A}$) under a condition on $\{\bar{x}_l: l \in \mathds{N}\}$, dependent on $k$. This condition is satisfied for all $k$ by the equidistant arrangement mentioned in ($\mathcal{A}$).   
\end{itemize}
In the sequel, the just presented agenda is called a \emph{staggering domino scenario}. Its validity is readily consistent with the observations based on the simulation results plotted in Fig. \ref{fig:ratios}.      
For $m=\mathcal{M}_1=1$, this scenario works for any initial arrangement: there occurs just one round of collisions, after which $u_0(t) = u_1(t) =0$, $u_2 (t) =1$. The last collision of the round (and the first at the same time) occurs at $x_2(t) =\bar{x}_2< \bar{x}_3$. In what follows, we show that it works for all $k>1$ and obtain the conditions mentioned in item ($\mathcal{B}$).

\subsection{The problem of velocities}

Here, we turn to calculating the velocities and studying the possibility that item ($\mathcal{A}$1), i.e., $u_{2l}(t) = u_{2l+1}(t) =0$, $u_{2l+2} (t) =1$, holds at a certain value of $\theta$.  To this end, we first study the collisions as if there were no other particles, except for those in the triple $0,1,2$. In this case, only their velocities matter, and time and space variables can be omitted at this stage. Since the velocities between collisions remain constant, it is more convenient for us to use their values updated at each collision round.      
For $k\in \mathds{N}_0$ and $l=0,1,2$, let $v_{l,k}$ be the corresponding velocity after round $k$. 
The condition for the next round to start is $v_{0,k}>0$. If this is the case, according to \eqref{1a} and \eqref{1}, the velocities after the next round are 
\begin{equation*}
\left(\begin{array}{l} v_{0,k+1}\\ v_{1,k+1} \\v_{2,k+1} \end{array} \right) = \left(\begin{array}{ccc} 
     1 \quad & 0 \quad & 0  \\
    0 \quad & - \theta \quad &1+\theta \\ 0 \quad & 1- \theta \quad &\theta
\end{array}\right) \left(\begin{array}{ccc} 
     \theta
    \quad & 1-\theta \quad & 0  \\
    1+\theta \quad & - \theta \quad & 0 \\ 0 \quad & 0 \quad & 1
\end{array}\right)
\left(\begin{array}{l} v_{0,k}\\ v_{1,k} \\v_{2,k} \end{array} \right) ,
\end{equation*}
which we abbreviate to the following formula
\begin{equation}
\label{3a}
V_{k+1} = A_1 A_0 V_k, \qquad V_k := \left(\begin{array}{l} v_{0,k}\\ v_{1,k} \\v_{2,k} \end{array} \right) ,
\end{equation}
where  $A_0$  and $A_1$ describe collisions $0\leftrightarrow 1$ and $1\leftrightarrow 2$, respectively.  
Then
\begin{equation}
\label{4}
 A:= A_1A_0= \left( \begin{array}{ccc}
    \theta \quad & 1-\theta \quad & 0   \\
   -\theta (1+\theta) \quad & \theta^2  & 1+\theta \\ 1-\theta^2 \quad & -\theta(1-\theta) \quad & \theta 
\end{array} \right) 
\end{equation}
describes the complete round of collisions. Note, however, that the collision $1\leftrightarrow 2$ in round $k+1$ 
may not occur if $v_{0,k+1}<0$.  
By \eqref{3a} and \eqref{4}, we get $v_{0,1} = \theta$, $v_{1,1} = -\theta(1+\theta)$, $v_{2,1} = 1-\theta^2$. 
For $\theta >0$, see \eqref{1}, the beginning of the second round is possible. After the collision $0\leftrightarrow 1$, one gets
\[
\left(\begin{array}{l} v_{0,2}\\ v'_{1,1}  \end{array} \right) =  \left(\begin{array}{ccc} 
     \theta
    \quad & 1-\theta   \\
    1+\theta \quad & - \theta 
\end{array}\right)
\left(\begin{array}{l} \qquad \theta \\ -\theta(1+\theta) \end{array} \right),
\]
and thus the corresponding velocities are $v_{0,2}= \theta^2 - \theta (1-\theta^2) =: \theta \phi_2(\theta)$,
$v'_{1,1} = \theta (1+\theta)^2$, $v_{2,1}= 1-\theta^2$. This means that the necessary condition for $1\leftrightarrow 2$ to occur is $v'_{1,1} > v_{2,1}$, which takes the form $\theta^2 + 2 \theta -1 >0$. If it holds, then the velocities after round 2 are obtained by applying $A$ to $V_1$, which yields
\begin{gather}
\label{M4a}
v_{0,2} = \theta \phi_2(\theta), \quad v_{1,2} = -(1+\theta)^2 \phi_2(\theta), \quad v_{2,2} =1 - \phi_2^2(\theta), \\[.2cm] \nonumber \phi_2(\theta) = \theta^2 + \theta -1.
\end{gather}
For $\theta \leq \theta_2:= (\sqrt{5}-1)/2$, which is the positive root of $\phi_2$, see \eqref{M4a}, we have $v_{0,2}\leq 0$. At the same time, the aforementioned condition $\theta^2 + 2 \theta -1 >0$ is verified for $\theta \geq \theta_2$. That is, $v_{0,k}\geq 0$ guarantees that both collisions in round $k$ occur.

By \eqref{M4a}, each $v_{l,k}$ is a polynomial in $\theta$. Keeping in mind the condition for the next round to occur, we define
\begin{equation}
\label{5}
\Theta_k = \{\theta \in [0,1): v_{0,k} >0\}.
\end{equation}
Then $\Theta_1 =(0,1)$ and 
$\Theta_2 = (\theta_2, 1)$. For $\theta=\theta_2$, i.e., for 
\begin{equation}
\label{M2}
m=\mathcal{M}_2:=2 +\sqrt{5},
\end{equation}
one has $v_{0,2} = v_{1,2}=0$ and $v_{2,2}=1$; see \eqref{M4a}. This yields the realization of item 
($\mathcal{A}$1) of the aforementioned scenario for $k=2$. 
In the Appendix \ref{A_I}, we present the calculations showing that item ($\mathcal{A}$1) holds for any $k\in \mathds{N}$ whenever
\begin{equation}
\label{M33}
m=\mathcal{M}_k := \frac{1}{2}\left[\tan^2 \frac{k\pi}{2k+1}-1\right] = \cot \frac{\pi}{2(2k+1)}\cot \frac{\pi}{2k+1}\, .
\end{equation}
Moreover, for $1\leq s \leq k$ and $m=\mathcal{M}_k$, the particle velocities after round $s$ are
\begin{gather}
\label{M33a}
v_{0,s} = - q_{2s-1} (\omega_k) q_{2s+1} (\omega_k), \qquad v_{1,s} = (2-\omega_k) q_{2s} (\omega_k) q_{2s+1} (\omega_k), \\ \nonumber v_{2,s} =  1- [q_{2s+1} (\omega_k)]^2 , \qquad \omega_k =1-\theta_k= \frac{2}{2\mathcal{M}_k +1} = \frac{2}{\tan^2\frac{k\pi}{2k +1}},
\end{gather}
where
\begin{gather}
\label{M33b}
q_{2s} (\omega) = \frac{\sin 2s \alpha}{\sqrt{\omega}\sin \alpha}, \qquad q_{2s+1} (\omega) = \frac{\sin(2s+1)\alpha }{\sin\alpha}, \qquad \alpha = \arctan \sqrt{\frac{4}{\omega}-1}.
\end{gather}
Note that
\begin{equation}
\label{M33c}
\alpha_k = \arctan\sqrt{\frac{4}{\omega_k}-1} = \arctan\sqrt{2\mathcal{M}_k +1} = \frac{k\pi}{2k+1},  
\end{equation}
see \eqref{M33} and \eqref{M33a}. Then by \eqref{M33a}, \eqref{M33b}, and \eqref{M33c} it follows that: (i) $v_{0,s}>0$ for $s\leq k-1$; (ii) $v_{0,k}= v_{1,k}=0$ and $v_{2,k}=1$.   

Formulas \eqref{M33} -- \eqref{M33c} provide the mathematical basis of the results of this paper.

\subsection{Points and times of collisions in the first triplet}

Now we turn to proving that items ($\mathcal{A}$2) and ($\mathcal{B}$) of the aforementioned scenario hold as well for the triplet $0,1,2$. Recall that $x_l (t)$, $l=0,1,2$ denote the positions of the particles at time $t$. To cover part ($\mathcal{B}$), we will deal with the general initial arrangement $x_l (t) = \bar{x}_l$.

For $s\geq 1$, let $\xi_s$, $\eta_s$, $\tau_s$, and $t_s$ denote the positions and the moments of time of the collisions $0\leftrightarrow 1$ and $1\leftrightarrow 2$ in round $s$, respectively. Since between collisions the particles move with constant velocities, it is convenient for us to deal with their positions, updated after each round, as we did in the previous subsection. We denote them    
$x_{l,s}$, $l=0,1,2$.  Then,   
for $t\in [t_s, \tau_{s+1}]$, the positions are
\begin{gather}
\label{D1}
x_{0,s}(t)= \xi_s + v_{0,s}(t-\tau_s), \quad x_{1,s}(t)= \eta_s + v_{1,s}(t-t_s), \\[.2cm] \nonumber x_{2,s}(t)= \eta_s + v_{2,s}(t-t_s), 
\end{gather}
with $v_{l,s}$ given in \eqref{M33a}. Then after the collision $0\leftrightarrow 1$ in round $s+1$, particles $0$ and $1$ move with velocities $v_{0,s+1}$ and $v_{1,s}' = (1+\theta)v_{0,s} - \theta v_{1,s}$. Note that $v_{1,0}'= 1+\theta$, $v'_{1,1} = \theta (1+\theta)^2$. Moreover, $v_{0,s}>0$ implies  $v'_{1,s}>0$, which by \eqref{2} yields $v'_{1,s}>1$, and hence 
\begin{eqnarray}
\label{D3a}
v'_{1,s}=(1+\theta) v_{0,s} - \theta v_{1,s} > v_{2,s},
\end{eqnarray}  
which indicates that the second collision in round $s+1$, i.e., $1\leftrightarrow 2$, will hold. 
For $t\in [\tau_{s+1}, t_{s+1}]$, i.e., before the second collision, the positions are
\begin{gather}
\label{D1z}
x_{0,s+1}(t)= \xi_{s+1} + v_{0,s+1}(t-\tau_{s+1}), \quad x_{1,s}(t)= \xi_{s+1} + v_{1,s}'(t-\tau_{s+1}), \\[.2cm] \nonumber x_{2,s}(t)= \eta_s + v_{2,s}(t-t_s). 
\end{gather}
Now, the updated values of the parameters $\xi_s$, $\eta_s$, $\tau_s$, and $t_s$ are to be found from the following four equations
\begin{gather}
 \label{D1y}
 \xi_{s+1} = x_{0,s} (\tau_{s+1}) = x_{1,s} (\tau_{s+1}), \qquad \eta_{s+1} = x_{1,s} (t_{s+1}) = x_{2,s} (t_{s+1}),
\end{gather}
from which by \eqref{D1} and \eqref{D1z} we obtain
\begin{gather}
\label{D1x}
\xi_s + v_{0,s} (\tau_{s+1} - \tau_s) = \eta_s + v_{1,s} (\tau_{s+1} - t_s), \qquad \xi_{s+1} + v_{1,s}' (t_{s+1}-\tau_{s+1}) = \eta_s + v_{2,s} (t_{s+1} - t_s).  
\end{gather}
 The initial values of the parameters in question are, cf. \eqref{M4a},
\begin{gather}
\label{D1a}
\tau_1=\bar{x}_1, \quad \xi_1 = \bar{x}_1, \quad t_1= \bar{x}_1+ \frac{\bar{x}_2 - \bar{x}_1}{1+\theta}, \quad \eta_1 =\bar{x}_2, \\[.2cm] \nonumber
\tau_2 = \bar{x}_1+ \frac{1+\theta}{\theta(2+\theta)}(\bar{x}_2 - \bar{x}_1), \quad \xi_2 = \bar{x}_1+ \frac{1+\theta}{2+\theta}(\bar{x}_2-\bar{x}_1), \\[.2cm] \nonumber t_2 = \bar{x}_1+ \frac{\theta(2+\theta)}{(1+\theta)(\theta^2 +2\theta-1)} (\bar{x}_2 - \bar{x}_1), \quad \eta_2 = \bar{x}_2 + \frac{1-\theta}{\theta^2+2\theta-1}(\bar{x}_2 - \bar{x}_1),
\end{gather}
where $\theta$, see \eqref{1}, will eventually be set to $\theta=\theta_k = (\mathcal{M}_k -1)/(\mathcal{M}_k +1)$. 
For $\theta = \theta_2$, see \eqref{M2}, we have $\theta_2^2 + \theta_2 -1=0$, and then by \eqref{D1a}, it follows that 
\begin{gather}
\label{D20}
\tau_2 = \bar{x}_2, \qquad \xi_2 = \bar{x}_1+ \theta_2 (\bar{x}_2- \bar{x}_1), \\ \nonumber
t_2 = x_1 + \frac{1}{\theta_2} (\bar{x}_2 -
\bar{x}_1) = \bar{x}_2 + \theta_2 (\bar{x}_2 -
\bar{x}_1), \quad \eta_2 = \bar{x}_2 + \theta_2 (\bar{x}_2 -
\bar{x}_1). 
\end{gather}
Then the condition $\eta_2 < \bar{x}_3$ mentioned in item ($\mathcal{B}$) of the staggering domino scenario is
\begin{eqnarray}
\label{D21}
\theta_2 (\bar{x}_2 -
\bar{x}_1) < \bar{x}_3 -
\bar{x}_2.
\end{eqnarray}
It is clearly satisfied for $\bar{x}_l=l$, as well as for any other equidistant arrangement.

After due calculations, we obtain from the equations in \eqref{D1y}, \eqref{D1x} the following recurrences  
\begin{eqnarray}
\label{D2}
\tau_{s+1} & = & \frac{v_{0,s}\tau_s}{v_{0,s}- v_{1,s}} - \frac{v_{1,s}t_s}{v_{0,s}- v_{1,s}} - \frac{\xi_{s}}{v_{0,s}- v_{1,s}} + \frac{\eta_{s}}{v_{0,s}- v_{1,s}}, \\[.2cm] \nonumber
\xi_{s+1 }& = & \frac{v_{0,s}v_{1,s}\tau_s}{v_{0,s}- v_{1,s}} - \frac{v_{0,s}v_{1,s}t_s}{v_{0,s}- v_{1,s}} - \frac{v_{1,s}\xi_s}{v_{0,s}- v_{1,s}} + \frac{v_{0,s}\eta_s}{v_{0,s}- v_{1,s}},
\end{eqnarray}
as well as
\begin{gather}
\label{D3b}
t_{s+1} = \frac{v_{0,s}(v'_{1,s}- v_{1,s})\tau_s}{(v'_{1,s}- v_{2,s})(v_{0,s}- v_{1,s})} - \frac{v_{1,s}(v'_{1,s}- v_{0,s})t_s}{(v'_{1,s}- v_{2,s})(v_{0,s}- v_{1,s})} \\[.2cm] \nonumber - \frac{v_{2,s}t_s}{v'_{1,s}- v_{2,s}} 
- \frac{(v'_{1,s}- v_{1,s})\xi_s}{(v'_{1,s}- v_{2,s})(v_{0,s}- v_{1,s})} \\[.2cm] \nonumber+ \frac{(v'_{1,s}- v_{0,s})\eta_k}{(v'_{1,s}- v_{2,s})(v_{0,s}- v_{1,s})}+ \frac{\eta_s}{v'_{1,s}- v_{2,s}}, 
\end{gather}
and 
\begin{gather}
\label{D3c}
\eta_{s+1} = \frac{v_{0,s}v_{2,s}(v'_{1,s}- v_{1,s})\tau_s}{(v'_{1,s}- v_{2,s})(v_{0,s}- v_{1,s})} - \frac{v_{1,s}v_{2,s}(v'_{1,s}- v_{0,s})t_s}{(v'_{1,s}- v_{2,s})(v_{0,s}- v_{1,s})} \\[.2cm] \nonumber - \frac{v'_{1,s}v_{2,s}t_s}{v'_{1,s}- v_{2,s}} 
- \frac{v_{2,s}(v'_{1,s}- v_{1,s})\xi_s}{(v'_{1,s}- v_{2,s})(v_{0,s}- v_{1,s})} \\[.2cm] \nonumber+ \frac{v_{2,s}(v'_{1,s}- v_{0,s})\eta_s}{(v'_{1,s}- v_{2,s})(v_{0,s}- v_{1,s})}+ \frac{v'_{1,s}\eta_s}{v'_{1,s}- v_{2,s}}.
\end{gather}
In the Appendix \ref{A_II}, the recurrences in \eqref{D2}, \eqref{D3b}, \eqref{D3c} are solved by means of the expressions for the velocities in \eqref{M33a}. The result is as follows. For $m=\mathcal{M}_k$, $k\in \mathds{N}$, the relevant parameters of the last round, i.e., for $s=k$, are: 
\begin{equation}
\label{D3w}
\xi_k = \bar{x}_1 +\theta_k(\bar{x}_2-\bar{x}_1), \quad \tau_k = \bar{x}_2 + \frac{\theta^2_k + \theta_k -1}{1+\theta_k}(\bar{x}_2-\bar{x}_1), \quad \eta_k = t_k= \bar{x}_2+\theta_k(\bar{x}_2-\bar{x}_1),
\end{equation}
cf. \eqref{D20}.
Recall that $\theta_k = (\mathcal{M}_k -1)/(\mathcal{M}_k +1)$, see \eqref{1}. Then the extension of \eqref{D21} to all $k$ takes the form
\begin{equation}
\label{D22}    
\theta_k (\bar{x}_2 -
\bar{x}_1) < \bar{x}_3 -
\bar{x}_2, \qquad k \geq 2.
\end{equation}
Thus, both items ($\mathcal{A}$) and ($\mathcal{B}$) of the staggering domino scenario hold for the triplet $0,1,2$.   

\subsection{Collisions in the entire system}

First, we consider the equidistant initial arrangement as in item ($\mathcal{A}$) of the staggering domino scenario. 
For any $k\geq 2$, for $m=\mathcal{M}_k$ and $t\geq t_k=2+\theta_k$, see \eqref{D3w}, particle $2$ moves from point $\eta_k = 2 +\theta_k$ to point $3$ with velocity $1$, where at $t=3$ it collides with particle $3$. This is the first collision in the triplet $2,3,4$. Since the particle masses in this triplet are the same as in the previous one, the same rules apply. Hence, this new triplet undergoes $k$ rounds of mutual collisions as in the case of the triplet $0,1,2$. The only difference is that the motion with a unit velocity now lasts $3-t_k= 1-\theta_k$, unlike the initial motion that lasts a unit time interval. At the same time, the first collision in the triplet $2,3,4$ occurs at point and time $3$, as if particle 2 started at time and point 0 and moved with a unit velocity without colliding with particles 1 and 2. The last collision in the triplet $2,3,4$ occurs at point and time $4+\theta_k$, as if particles 0,1, and 2 were taken out of the system and particle 2 started from point 0. This resembles the usual domino dynamics. 

To describe the motion of the entire system, let us number the triplets by $l \in \mathds{N}_0$ according to the corresponding leftmost particle. That is, triplet $l$ consists of particles $2l, 2l+1, 2l+2$. Let $t_{l,k}$ and $\eta_{l,k}$ be the time and position of the last collision in triplet $l$. For example, $t_{0,k} = \eta_{0,k}= 2+\theta_k$. Then the first collision in triplet $l$ occurs at point and time $2l +1$, and $t_{l,k} = \eta_{l,k}= 2(l +1)+\theta_k< 2l +3$. This proves that both parts of item ($\mathcal{A}$) of the staggering domino scenario hold. Moreover, according to the arguments just presented, the shock front position satisfies 
\begin{equation}
 \label{Rnu}
 \mathcal{R}(2l+1) = 2l +1,
\end{equation}
i.e., it moves with an average unit velocity. At the same time, the actual velocity 
of the shock front ranges from $v_{2,1} = 1- [q_3 (\omega_k)]^2 = 1 - \theta_k^2$, see \eqref{M33a}, to $v'_{1,1} = 1 + \theta_k$. For large $k$, the former velocity is close to zero, while the latter is nearly two.  

For the general initial arrangement, the condition mentioned in item ($\mathcal{B}$) is, cf. \eqref{D22}, 
\begin{equation}
\label{D24}
\theta_k (\bar{x}_{2l} -
\bar{x}_{2l-1}) < \bar{x}_{2l+1} -
\bar{x}_{2l}, \qquad l \geq 1.
\end{equation}
Since $\theta_{k-1} < \theta_k$ and $\theta_k \to 1$ as $k\to +\infty$, the validity of \eqref{D24} for a given $k$ implies its validity for all $k'< k$, but not for $k''>k$. For $\bar{x}_{2l} -
\bar{x}_{2l-1} \leq \bar{x}_{2l+1} -
\bar{x}_{2l}$, the condition in \eqref{D24} is satisfied for all $k$.  Now, let us allow each $\bar{x}_l$, $l\geq 1$, to deviate from $l$, i.e., for a fixed $k$, find $\varepsilon_k$ such that \eqref{D24}  
holds whenever $|\bar{x}_l-l|< \varepsilon_k$. A simple calculation yields $\varepsilon_k = 1/2\mathcal{M}_k$.
That is, such deviations from the equidistant arrangement $\bar{x}_l =l$, $l\in \mathds{N}_0$ do not affect the staggering domino dynamics.

Now we turn to considering general initial arrangements. Assuming the validity of \eqref{D24}, by \eqref{D3w} we obtain $x_2(t) = \eta_k + (t-t_k) = t$, as is for $t>t_k$. That is, the first collision $2\leftrightarrow 3$ occurs at $t=\bar{x}_3$. Similarly to the equidistant arrangement, we can interpret this as if particle 2 started at time and point 0 and moved at unit velocity without colliding with particles 1 and 2. Then the triplet $2,3,4$ performs $k$ rounds of collisions, and the corresponding parameters, cf. \eqref{D3w}, take the form
$\xi_{1,k} = \bar{x}_3 + \theta_k (\bar{x}_4-\bar{x}_3)$, $\eta_{1,k} = t_{1,k}= \bar{x}_4 + \theta_k (\bar{x}_4-\bar{x}_3)$. Repeating these arguments, we conclude that the same staggering domino dynamics occur with each triplet 
 $2l, 2l+1, 2l + 2$. The corresponding parameters take the following form
 \begin{gather}
 \label{D25}
 \xi_{l,k} = \bar{x}_{2l+1} + \theta_k (\bar{x}_{2l+2}-\bar{x}_{2l+1}), \qquad \tau_{l,k} = \bar{x}_{2l+2} + \frac{\theta^2_k + \theta_k -1}{1+\theta_k} (\bar{x}_{2l+2}- \bar{x}_{2l+1}), \\ \nonumber \eta_{l,k} = t_{l,k} = \bar{x}_{2l+2} + \theta_k (\bar{x}_{2l+2}-\bar{x}_{2l+1}), \qquad \mathcal{R}(\bar{x}_{2l+1}) = \bar{x}_{2l+1}.
 \end{gather}
 The latter implies that the shock front moves in a ballistic way.

\section{Conclusions and Outlook}\label{VI}

In this paper, we study a system of point particles with masses $m,1,m,1, \dots$ located in $\mathds{R}_{+} = [0,+\infty)$. The leftmost particle starts moving to the right, initiating the dynamics of the whole system through elastic collisions. For $m=2,3,10$ and the equidistant initial positions of the particles, our results presented in Fig \ref{fig:powers} indicate that the dynamics is characterized by a blast wave propagating to the right in a hydrodynamic way and splashes moving ballistically to the left. These results are consistent with those reported in \cite{Chakraborti22} for the same model with $m=2$ and random initial positions. At the same time, the key result of this paper is as follows. We found a countably infinite set $\{\mathcal{M}_k: n \in \mathds{N}\}$, see \eqref{M33}, such that for $m=\mathcal{M}_k$ the system dynamics may follow the staggering domino scenario presented in section \ref{V} and outlined in Fig. \ref{fig:triplets}.
The condition for this scenario to be realized for a given $m=\mathcal{M}_k$ is that the initial positions satisfy \eqref{D24}. It is realized for all $k\geq 2$ under the condition $\bar{x}_{l+1} - \bar{x}_{l} \geq \bar{x}_{l} - \bar{x}_{l-1}$, $l\geq 2$, which is satisfied by the equidistant arrangements. In \cite{Holovatch25b}, this scenario was originally proposed as a hypothesis and then verified by symbolic computation for $k\leq700$ and equidistant initial arrangements. In this paper, we prove this hypothesis for all $k$, i.e., for the values of $m$ that form a countably infinite set $\{\mathcal{M}_k: n \in \mathds{N}\}$. Our results cover general initial arrangements satisfying \eqref{D24}.

To gauge the significance of our results, we note that they imply that the staggering domino dynamics may also be observed in a broader class of systems. Indeed, 
let us modify the boundary conditions imposed on the system we study in the following ways: (a) the negative half-line $(-\infty, 0)$ is occupied by alternating mass particles of the same kind; consequently, the aforementioned splatter can enter this region
and undergo collisions with these particles, cf. \cite{Antal08}; (b) there exists an elastic reflecting screen at $x=0$ which redirects the splatter.
For \(m \neq \mathcal{M}_k\), the splatter exists and, due to the boundary conditions in (a) and (b), it bounces back into $\mathds{R}_{+}$, essentially affecting the motion of the shock front. By our results, for \(m = \mathcal{M}_k\), the splatter is absent, and thus the shock front continues its staggering domino motion also if boundary conditions (a) or (b) are applied.

As an important aspect of our research, we also note that it covers all initial arrangements that satisfy the condition in \eqref{D24}. In particular, random shifts of the initial positions $\bar{x}_l$ from integer $l$ satisfying $|\bar{x}_l-l|< \varepsilon_k$ do not affect the staggering domino effect for a given $m=\mathcal{M}_k$, whenever $\varepsilon_k\leq 1/2\mathcal{M}_k$. Another possibility of using \eqref{D24} consists in the following. Assume that $X_l := \bar{x}_{l+1}- \bar{x}_{l}$, $l\geq 1$, are random and that they are independent and identically distributed. Then by means of \eqref{D24}, one can try to estimate the probability that the staggering domino shock front reaches a given point $x\in \mathds{R}_{+}$.  

There is one further significant aspect of the current research worth mentioning. The hydrodynamic description summarized in Table \ref{tab1} was originally derived in 
\cite{Chakraborti22} by analyzing the scaling properties of potential solutions to the corresponding Euler equation. However, for one-dimensional gas models, solving the Euler equation can also yield $\delta$-shock waves that propagate at constant velocities; see \cite[Chapter II]{Sheng99}. Consequently, our results may be viewed as a microscopic counterpart to this specific theoretical framework.

It seems natural to expect that the staggering domino-like motion is possible not only in a one-dimensional cold gas model with two alternating masses. Indeed, we found that the same effect occurs for the trinomial distribution of masses \((m_x, m_y, m_z)\), with \(m_x = \mathcal{M}_3 = \cot \frac{\pi}{14} \cot \frac{\pi}{7}\), \(m_y = (1 + 3m_x)/(m_x - 1)\), and \(m_z=1\), albeit under even stricter constraints than for equidistant initial positions.

By examining the transition from discrete particle dynamics to a continuum description, we identify the specific conditions under which the blast wave preserves its scaling properties. In particular, our results demonstrate that the interplay between the mass ratio and initial particle locations significantly influences the propagation of the shock front. This naturally raises the question of how the system passes from the dynamics of an AHP cold gas with equidistant particles to the behavior observed with random arrangements. This transition, along with the ergodic properties of the system and the role of random initial arrangements, will be addressed in our future work.

\appendix

\section{Dealing with velocities}\label{A_I}

The mathematical basis of the material in these appendices amounts to elementary linear algebra and trigonometry.
First, we derive formulas \eqref{M33} -- \eqref{M33c}. Then, in the Appendix \ref{A_II}, we solve the recursion relations in \eqref{D2} -- \eqref{D3c}.

The first line of \eqref{M1a} allows one to exclude $v_{2,k}$ without losing the linearity of 
the recursion relation in \eqref{3a}. To this end, we introduce
\begin{equation}
\label{M6}
w_{0,k} = v_{0,k}, \qquad w_{1,k} = \frac{v_{1,k}}{\theta+1},  
\end{equation}
and then get
\begin{equation}
\label{M7}
v_{2,k} = 1- w_{0,k} - (1-\theta) w_{1,k}.
\end{equation}
Then by \eqref{4} and \eqref{M6}, \eqref{M7}, we obtain the following recursion relation
\begin{gather}
\label{M8}
W_k = B W_{k-1} + P_0, \qquad W_k = \left(\begin{array}{cc} w_{0,k} \\ w_{1,k}    
\end{array} \right), \quad P_0 =  \left(\begin{array}{cc} 0 \\ 1    
\end{array} \right),
\end{gather}
with
\begin{gather}
\label{M8B}
 B = \left( \begin{array}{cc}
   \theta \quad & 1-\theta^2  \\
  -(1+\theta) \quad   & \phi_2(\theta) 
\end{array}\right),   
\end{gather} 
where $\phi_2$ is as in \eqref{M4a}. The recursion relation in \eqref{M8} is subject to the initial condition $w_{0,0}=1$, $w_{1,0}=0$, which yields 
$w_{0,1}=\theta$, $w_{1,1}= - \theta$, and, cf. \eqref{M4a}, 
\begin{equation}
 \label{M9}
 w_{0,2} = \theta \phi_2(\theta), \qquad w_{1,2} =  - (\theta^3 + 2 \theta^2 -1) = -(1+\theta )\phi_2(\theta). 
\end{equation}
The solution of the recursion in \eqref{M8} is
\begin{eqnarray}
\label{M10}
W_k & = & B^k W_0 + (B^{k-1} + \cdots + B + I)P_0 \\[.2cm] \nonumber & = & B^k W_0 + (B^k-I) P_1 = B^k (W_0+P_1) - P_1, \\[.2cm] \nonumber P_1 & = & (B-I)^{-1} P_0  = - \frac{1-\theta}{3 - 2\theta - \theta^2}   \left(\begin{array}{cc}  
 1+\theta \\ 1     
\end{array} \right),
\end{eqnarray}
where $I$ is the unit $2\times 2$ matrix. In view of the latter, we set
\begin{equation}
 \label{M11}
 W_0 + P_1 =    \frac{1-\theta}{3 - 2\theta - \theta^2} Q_0 = \frac{1}{\theta +3} Q_0, \quad Q_0 :=  \left(\begin{array}{cc}  
 2 \\ -1     
\end{array} \right). 
\end{equation}
By \eqref{M10} and \eqref{M11} we then get
\begin{gather}
\label{M12}
W_k =  \frac{1}{\theta +3} \left[ B^k Q_0 + \left(\begin{array}{cc}  
 1+\theta \\ 1     
\end{array} \right) \right].
\end{gather}
Now we turn to studying the powers of $B$ given in \eqref{M8B}. For $\omega \in (0,1)$, set
\begin{equation}
\label{C}
C(\omega) = \left(\begin{array}{cc} 0 \quad
     & -\omega  \\
  1 \quad    &\omega 
\end{array} \right).
\end{equation}
Then
\begin{eqnarray}
\label{M13}
C^2 (\omega) = \omega[C(\omega) - I], \quad {\rm and} \quad
B = \theta I - (1+\theta) C(1-\theta).
\end{eqnarray}
For $\omega = 1-\theta$, by \eqref{M13} we get
\begin{gather}
\label{M15}
B = (1-\omega) I - (2-\omega) C(\omega) = \omega[C(\omega) - I] +I - 2C(\omega) \\[.2cm] \nonumber = C^2(\omega) - 2C(\omega) +I = [C(\omega) - I]^2 = \frac{1}{\omega^2} C^4(\omega).
\end{gather}
Then the problem is reduced to studying the powers of $C(\omega)$. By \eqref{M13} one can guess that 
\begin{eqnarray}
\label{M16}
C^k (\omega) = p_k(\omega) C(\omega) - \omega p_{k-1}(\omega) I, \quad p_2(\omega)=\omega, \quad  p_1(\omega) \equiv 1, \quad p_0(\omega) \equiv 0,
\end{eqnarray}
which indeed holds if the polynomials $p_k$ satisfy
\begin{eqnarray}
  \label{M17}
  p_{k+1}(\omega) =  \omega[p_{k}(\omega)- p_{k-1}(\omega)], \qquad k\geq 1.
\end{eqnarray}
Starting from $p_0$, $p_1$ given in \eqref{M16}, by \eqref{M17} one obtains
\begin{gather}
\label{M18}
    p_{2}(\omega) = \omega, \quad p_{3}(\omega) = \omega^2 - \omega, \quad p_{4}(\omega) = \omega^3 - 2 \omega^2, \\[.2cm] \nonumber p_{5}(\omega) = \omega^4 - 3 \omega^3 + \omega^2, \quad p_{6}(\omega) = \omega^5 - 4 \omega^4 +3 \omega^3,
\end{gather}
and so on. The recurrence in \eqref{M17} generates the Fibonacci polynomials; see, e.g., \cite{Swamy}. Therefore, it is solved by an analog of Binet's formula as follows. Let $a$ and $b$ be two (complex) roots of the polynomial $\lambda^2 - \omega \lambda + \omega$, that is,
\begin{equation}
 \label{M19}
 a = \frac{1}{2}[\omega + i \sqrt{4\omega - \omega^2}], \quad b = \frac{1}{2}[\omega - i \sqrt{4\omega - \omega^2}]. 
\end{equation}
Then the solution of \eqref{M17} is
\begin{equation}
\label{M20}
p_k (\omega) = \frac{a^k-b^k}{a-b} = \omega^{(k-1)/2} \frac{\sin k\alpha}{\sin \alpha}, \qquad \alpha = \arctan \sqrt{\frac{4}{\omega} -1}.
\end{equation}
Recall that $\omega = 1-\theta \in (0,1)$, and 
\begin{equation}
\label{M21}
\cos \alpha = \frac{\sqrt{\omega}}{2}, \qquad 
\sin \alpha= \frac{1}{2}\sqrt{4 - \omega}.
\end{equation}
By \eqref{M21} and \eqref{M20}, one easily gets $p_k$ as given in \eqref{M18}. In the sequel, we will use the following formulas, which can be obtained from \eqref{M21}
\begin{gather}
\label{M22}
\cos 2 \alpha = - \frac{2-\omega}{2}, \qquad \sin 2\alpha = \frac{\sqrt{\omega}}{2} \sqrt{4-\omega}, \\[.2cm] \nonumber \sin 3\alpha = -\frac{1}{2}(1-\omega) \sqrt{4-\omega}, 
\end{gather}
and so on. By \eqref{M12}, \eqref{M15} and \eqref{M16} we then get
\begin{eqnarray}
\label{M23}
w_{0,k} & = & \frac{1}{4-\omega}\left[ 2-\omega - \frac{2}{\omega^{2k-1}} p_{4k-1} (\omega) + \frac{1}{\omega^{2k-1}} p_{4k} (\omega) \right], \\[.2cm] \nonumber w_{1,k} & = & \frac{1}{4-\omega}\left[ 1 + \frac{1}{\omega^{2k-1}} p_{4k-1} (\omega) + \frac{2-\omega}{\omega^{2k}} p_{4k} (\omega) \right],
\end{eqnarray}
that holds for all $k\in\mathds{N}$.
Now we use in \eqref{M23} $p_k$ as given in \eqref{M20} and the formulas in \eqref{M22} to get the following
\begin{eqnarray}
\label{M24}
v_{0,k} &  = & w_{0,k}=\frac{1}{4-\omega}\left[ 2-\omega - 2 \frac{\sin(4k-1)\alpha}{\sin \alpha} + \sqrt{\omega} \frac{\sin4k\alpha}{\sin \alpha} \right] \\[.2cm] \nonumber & = & \frac{2}{(4-\omega)\sin \alpha}\bigg{[}- \cos 2 \alpha \sin \alpha + 2 \cos \alpha \sin 4k \alpha - 2 \sin (4k-1)\alpha \bigg{]} \\[.2cm] \nonumber 
& = & \frac{2}{(4-\omega)\sin \alpha}\bigg{[}- \cos 2 \alpha \sin \alpha + \cos 4k\alpha \sin \alpha \bigg{]} \\[.2cm] \nonumber 
& = & -\frac{4}{(4-\omega)} \sin (2k+1) \alpha \sin (2k-1) \alpha= - \omega^{-(2k-1)}p_{2k-1}(\omega)p_{2k+1}(\omega).
\end{eqnarray}
By \eqref{M21}, \eqref{M22} applied in \eqref{M24}, one calculates  
\[
v_{0,1}=w_{0,1}= - \frac{4}{(4-\omega)} \sin 3 \alpha \sin  \alpha = 1- \omega = \theta,
\]
which agrees with \eqref{M2}. Now we use the second line of \eqref{M23} to get 
\begin{eqnarray}
\label{M25}
w_{1,k} &  = & \frac{1}{4-\omega}\left[ 1 +  \frac{\sin(4k-1)\alpha}{\sin \alpha} + \frac{2-\omega}{\sqrt{\omega}} \frac{\sin4k\alpha}{\sin \alpha} \right] \\[.2cm] \nonumber & = & \frac{1}{(4-\omega)\sin \alpha}\bigg{[} \sin  \alpha  +  \sin (4k -1)\alpha +  \frac{2-\omega}{\sqrt{\omega}} \sin 4k\alpha \bigg{]} \\[.2cm] \nonumber 
& = & \frac{2}{(4-\omega)\sqrt{\omega}\sin \alpha}\bigg{[} \cos  \alpha \sin \alpha + \sin (4k-1)\alpha \cos \alpha - \sin 4k\alpha \cos 2\alpha \bigg{]} \\[.2cm] \nonumber 
& = & \frac{2}{(4-\omega)\sqrt{\omega}\sin \alpha}\bigg{[} \cos  \alpha \sin \alpha  -  
\cos 4k\alpha \sin \alpha \cos \alpha  + \sin 4k\alpha \sin^2 \alpha \bigg{]}
\\[.2cm] \nonumber 
& = &
\frac{2}{(4-\omega)\sqrt{\omega}} \bigg{[} \cos \alpha - \cos 4k\alpha \cos \alpha +  \sin 4k \alpha \sin \alpha \bigg{]} \\[.2cm] \nonumber 
& = &
\frac{2}{(4-\omega)\sqrt{\omega}}  [\cos \alpha - \cos (4k+1) \alpha] = \frac{4}{(4-\omega)\sqrt{\omega}} \sin (2k+1) \alpha \sin 2k\alpha\\[.2cm] \nonumber 
& = & \omega^{-2k} p_{2k}(\omega)p_{2k+1}(\omega). 
\end{eqnarray}
Similarly as above, $w_{1,1} = \omega^{-2} p_{2}(\omega)p_{3}(\omega)= \omega-1 = - \theta$, see \eqref{M18}. In view of \eqref{M6}, \eqref{M7}, by \eqref{M25} it follows that
\begin{gather}
\label{M26}
v_{1,k} = \omega^{-2k}(2-\omega) p_{2k}(\omega)p_{2k+1}(\omega), \\[.2cm] \nonumber v_{2,k} = 1 - \omega^{-(2k-1)} p_{2k + 1} (\omega)[p_{2k} (\omega) - p_{2k-1} (\omega)] = 1- [\omega^{-k}p_{2k + 1} (\omega)]^2 . 
\end{gather}
A more convenient form of \eqref{M24}, \eqref{M26} employs the polynomials
\[
q_{2k} (\omega)= \omega^{-k}p_{2k} (\omega)= \frac{\sin 2k\alpha}{\sqrt{\omega}\sin \alpha}, \quad q_{2k+1} (\omega)= \omega^{-k}p_{2k+1} (\omega) = \frac{\sin (2k+1)\alpha}{\sin \alpha}, \] 
which yields
\begin{gather}
\label{M27}
v_{0,k}= -  q_{2k-1} (\omega)q_{2k+1} (\omega), \quad v_{1,k}=(2 - \omega)  q_{2k} (\omega)q_{2k+1} (\omega), 
\\[.2cm] \nonumber  v_{2,k}= 1 -  [q_{2k+1} (\omega)]^2.
\end{gather}
Recall that, see \eqref{M21} and \eqref{1},
\begin{equation}
\label{M27a}
\tan \alpha = \sqrt{\frac{4}{\omega}-1} = \sqrt{\frac{3+\theta}{1-\theta}} = \sqrt{2m + 1} \geq \sqrt{3}.
\end{equation}
Now, for a fixed $k\geq 2$, let us find $\alpha \in [\pi/3, \pi/2)$ (hence $m$, see \eqref{M27a}) such that the following holds
\begin{eqnarray}
\label{M28}
\sin (2k+1) \alpha =0, \qquad \sin (2s-1) \alpha \sin (2s+1) \alpha < 0, \quad {\rm for} \ {\rm all } \ 2\leq s \leq k-1.
\end{eqnarray}
 By \eqref{M27}, for such $\alpha$, 
\begin{equation}
\label{M29}
v_{0,k} = v_{1,k} = 0, \quad v_{2,k}=1, \quad {\rm and} \quad v_{0,s} >0, \quad {\rm for} \ {\rm all } \ 2\leq s\leq k-1,
\end{equation}
cf. item ($\mathcal{A}$) of the staggering domino scenario.
By the first condition in \eqref{M28}, it follows that
\begin{equation}
\label{M30}
\alpha = \frac{l \pi}{2k+1}, \qquad \frac{2k+1}{3} \leq  l \leq k.
\end{equation}
For $k = 2,3$, the only choice in \eqref{M30} is $l=k$. At the same time, one can take $l=k-1,k$ starting from $k=4$ and $l=k-2, k-1,k$ starting from $k=7$. In general, one can take $l\geq k-q$ for $k\geq 3q+1$. Let us show that the choice $l=k$ verifies also the second condition in \eqref{M28}. To this end, write
\begin{equation}
\label{M31}
\frac{(2s-1)k}{2k+1} = s - \frac{k+s}{2k+1}, \qquad \frac{(2s+1)k}{2k+1} = s + \frac{k-s}{2k+1}.
\end{equation}
Then, for $\alpha= k\pi/(2k+1)$, it follows that 
\begin{gather*}
\sin (2s\pm 1)\alpha = \sin \left( s\pi \pm \frac{k\mp s}{2k +1}\pi \right) =  \pm (-1)^s \sin\frac{(k\mp s)\pi}{2k+1},\\   \sin (2s- 1)\alpha \sin (2s+ 1)\alpha= - \sin\frac{(k+ s)\pi}{2k+1} \sin\frac{(k- s)\pi}{2k+1}<0,
\end{gather*}
which holds for $s$ satisfying \eqref{M29}.
Now we set $l = k-q$, $q\geq 1$ and take into account that $k \geq 3q+1$. Similarly to \eqref{M31}, we then write
\begin{gather}
 \label{M32}
 \frac{(2s-1)(k-q)}{2k+1} = s- \frac{s(2q+1) +k -q}{2k+1}=:s - \delta_{-}, \\[.2cm] \nonumber \frac{(2s+1)(k-q)}{2k+1} = s - \frac{s(2q+1) -k +q}{2k+1}=:s - \delta_{+},
\end{gather}
which yields
\begin{equation}
\label{M32z}
\varkappa(k,q,s):=\sin (2s- 1)\alpha \sin (2s+ 1)\alpha= \sin \delta_{+} \pi \sin\delta_{-}\pi.
\end{equation}
Let us show that, for each $q\geq 1$ and $k\geq 3q+1$, an integer $s\in [2, k-1]$ can be found such that $\varkappa(k,q,s)\geq 0$.  By \eqref{M32z}, this is equivalent to showing $0\leq \delta_{+} < \delta_{-}\leq 1$. The latter can be transformed into 
\[
\frac{k-q}{2q+1 } \leq s \leq \frac{k+q+1}{2q+1 }, 
\]
which readily holds with $s\geq 2$, since the above interval for $s$ is of unit length. Hence, it contains at least one integer for every $k\geq 3q+1$.
For instance, one takes $s=2$ for $k=3q+1$. By \eqref{M32z} and \eqref{M27},
 this implies that $v_{0,s}$ does not satisfy the second condition in \eqref{M29}. Thus, the only admissible choice in \eqref{M30} is $l=k$. By \eqref{M27a} we then get the validity of the expressions in \eqref{M33} -- \eqref{M33c}.

\section{Dealing with points and times of collisions}\label{A_II}

For a fixed $k\geq 2$, hence fixed $m=\mathcal{M}_k$ and $\theta_k$, $\alpha_k$, see \eqref{M33}, \eqref{M33c}, we solve the recursions in \eqref{D2} -- \eqref{D3c}. 
Our first aim is to exclude $\tau_s$ and $t_s$ from these recursions, i.e., to concentrate on $\xi_s$ and $\eta_s$.
By \eqref{D1} -- \eqref{D1x}, we get
\begin{gather}
\label{J2}
t_{s+1}-\tau_{s+1} = \frac{1}{v'_{1,s}} (\eta_{s+1}-\xi_{s+1}), \quad \tau_{s+1}-\tau_s = \frac{1}{v_{0,s}} (\xi_{s+1}-\xi_{s})  , \\ \nonumber t_{s+1}-t_s = \frac{1}{v_{2,s}} (\eta_{s+1} - \eta_s).
\end{gather}
Thereby, 
\begin{gather}
\label{J3}
\xi_{s+1}-\xi_{s} = \eta_s -\xi_s + v_{1,s}(\tau_{s+1} - \tau_s) - v_{1,s}(t_{s} - \tau_s) \\[.2cm] \nonumber = \eta_s -\xi_s + \frac{v_{1,s}}{v_{0,s}}(\xi_{s+1}-\xi_{s}) - \frac{v_{1,s}}{v'_{1,s-1}} (\eta_s -\xi_s).    
\end{gather}
The solution of the latter reads
\begin{gather}
\label{J4}
\xi_{s+1}-\xi_{s} = \frac{v_{0,s}(v'_{1,s-1}- v_{1,s})}{v'_{1,s-1}(v_{0,s}- v_{1,s})}(\eta_s -\xi_s).
\end{gather}
In a similar way, we have
\begin{gather*}
%\label{J5}
\eta_{s+1}-\eta_{s} = v_{2,s} (t_{s+1} - \tau_{s+1}) + v_{2,s} (\tau_{s+1} - \tau_{s}) - v_{2,s} (t_{s} - \tau_{s})  \\[.2cm] \nonumber = \frac{v_{2,s}}{v'_{1,s}} (\eta_{s+1}- \xi_{s+1}) + \frac{v_{2,s}}{v_{0,s}} (\xi_{s+1}- \xi_{s}) - \frac{v_{2,s}}{v'_{1,s-1}} (\eta_{s}- \xi_{s}) \\[.2cm] \nonumber = \frac{v_{2,s}}{v'_{1,s}} (\eta_{s+1}- \eta_{s}) - \frac{v_{2,s}}{v'_{1,s}} (\xi_{s+1}- \xi_{s}) + \frac{v_{2,s}}{v'_{1,s}} (\eta_{s}- \xi_{s})\\[.2cm] \nonumber
+ \frac{v_{2,s}}{v_{0,s}} (\xi_{s+1}- \xi_{s}) - \frac{v_{2,s}}{v'_{1,s-1}} (\eta_{s}- \xi_{s}).
\end{gather*}
By means of \eqref{J3}, we solve the latter and get
\begin{equation}
\label{J6}
\eta_{s+1}-\eta_{s} = \frac{v_{2,s}(v'_{1,s-1}- v_{0,s})(v'_{1,s}-v_{1,s})}{v'_{1,s-1} (v'_{1,s}-v_{2,s})(v_{0,s}-v_{1,s})}(\eta_{s}- \xi_{s}). 
\end{equation}
Next, we subtract \eqref{J3} from \eqref{J6} and obtain
\begin{equation}
\label{J7}
\eta_{s+1}-\xi_{s+1} = \frac{v'_{1,s}(v'_{1,s-1}- v_{0,s}) (v_{2,s}-v_{1,s})}{v'_{1,s-1}(v'_{1,s}- v_{2,s}) (v_{0,s}-v_{1,s})} (\eta_{s}-\xi_{s}),
\end{equation}
which is a recursion relation for $\eta_{s}-\xi_{s}$. We aim to solve it.
By \eqref{M2}, we have 
\begin{gather*}
v_{0,s+1}= \theta_k v_{0,s} + (1-\theta_k) v_{1,s}, \quad v'_{1,s}= (1-\theta_k) v_{0,s} - \theta_k v_{1,s}. \end{gather*}
By subtracting these equations, we obtain
\begin{equation}
\label{J8}
v_{0,s}- v_{1,s} = v'_{1,s} - v_{0,s+1},
\end{equation}
and similarly
\begin{equation}
\label{J9}
v_{2,s}- v_{1,s} = v'_{1,s-1} - v_{2,s-1}.
\end{equation}
Now by means of \eqref{J8}, \eqref{J9} we rewrite \eqref{J7} as follows
\begin{equation}
\label{J10}
\eta_{s+1}-\xi_{s+1} = \frac{v'_{1,s}(v'_{1,s-1}- v_{0,s}) (v'_{1,s-1}-v_{2,s-1})}{v'_{1,s-1} (v'_{1,s}-v_{0,s+1})(v'_{1,s}- v_{2,s})} (\eta_{s}-\xi_{s}) =: c_s (\eta_{s}-\xi_{s}), 
\end{equation}
where 
\begin{equation}
\label{J11}
c_s = \frac{v'_{1,s} \lambda_{s-1}}{v'_{1,s-1} \lambda_{s}}, \qquad \lambda_s = (v'_{1,s}-v_{0,s+1}) (v'_{1,s}-v_{2,s}).
\end{equation}
In particular, $\lambda_0 = 1+\theta_k$ and
\begin{equation}
\label{J12}
c_1 = \frac{1+\theta_k}{(2+\theta_k)(\theta_k^2 + 2\theta_k-1)}.
\end{equation}
Recall that we consider the case of $k\geq 2$, which means $\theta_k \geq \theta_2 =(\sqrt{5}-1)/2$. Hence, the denominator in \eqref{J12} is positive.
We then iterate \eqref{J10}, take into account \eqref{D1a} and \eqref{J11}, and thus obtain
\begin{equation}
\label{J13}
\eta_{s+1}-\xi_{s+1} =\frac{v'_{1,s}\lambda_0}{v'_{1,0}\lambda_s} (\eta_{1}-\xi_{1}) = \frac{v'_{1,s}(\bar{x}_2- \bar{x}_1)}{(v'_{1,s}-v_{0,s+1})(v'_{1,s}-v_{2,s})}.
\end{equation}
We use the latter in \eqref{J6}, which yields
\begin{equation}
 \label{J14}
 \eta_{s+1}-\eta_{s} = \frac{v_{2,s}(\bar{x}_2- \bar{x}_1)}{v_{0,s}- v_{1,s}}\left[\frac{1}{v_{2,s} - v_{1,s}} + \frac{1}{v_{2,s+1} - v_{1,s+1}}\right].
\end{equation}
Now we rewrite \eqref{M33a} in the form
\begin{eqnarray}
\label{J15}
v_{0,s}& = &  - \frac{\sin (2s-1)\alpha_k \sin (2s+1)\alpha_k}{\sin^2\alpha_k},\\[.2cm] \nonumber
v_{1,s}& = &  - \frac{\cos 2\alpha_k\sin 2s\alpha_k \sin (2s+1)\alpha_k}{\cos \alpha_k\sin^2\alpha_k}, \\[.2cm] \nonumber
v_{2,s}& = &  - \frac{\sin 2s\alpha_k \sin (2s+2)\alpha_k}{\sin^2\alpha_k}.
\end{eqnarray}
We use these formulas to calculate
\begin{eqnarray}
\label{J16}
v_{0,s} - v_{1,s} & = & \frac{\sin(4s+2)\alpha_k}{\sin 2 \alpha_k}, \\[.2cm] \nonumber
v_{2,s} - v_{1,s} & = & - \frac{\sin4s\alpha_k}{\sin 2 \alpha_k}, \\[.2cm] \nonumber 
\frac{v_{2,s}}{v_{0,s}- v_{1,s}} & = & - \frac{\cos \alpha_k\sin 2 s \alpha_k \sin (2s+2) \alpha_k }{\sin \alpha_k\sin (2s+1) \alpha_k \cos (2s+1)\alpha_k},
\end{eqnarray}
and also 
\begin{equation}
\label{J17}
\frac{1}{v_{2,s} - v_{1,s}}+ \frac{1}{v_{2,s+1} - v_{1,s+1}}
 =  - \frac{\sin 4\alpha_k \sin(4s+2)\alpha_k}{\sin 4s  \alpha_k \sin (4s+4) \alpha_k }.
\end{equation}
Next, we use \eqref{J16}, \eqref{J17} in \eqref{J14} and obtain
\begin{equation}
\label{J18}
\eta_{s+1}= \eta_s + (\bar{x}_2 - \bar{x}_1)\delta_s (\alpha_k):= \eta_s + \frac{(\bar{x}_2 - \bar{x}_1)\cos \alpha_k \sin 4\alpha_k}{2 \sin \alpha_k \cos 2s\alpha_k \cos (2s+2 ) \alpha_k},
\end{equation}
which finally yields, see \eqref{D1a},
\begin{eqnarray}
\label{J20}
\eta_s & = & \eta_1 + (\bar{x}_2 - \bar{x}_1)\sum_{l=1}^{s-1} \delta_l (\alpha_k)\\[.2cm] \nonumber & = & \bar{x}_2 + (\bar{x}_2 - \bar{x}_1)\sum_{l=1}^{s-1} \frac{\cos \alpha_k \sin 4\alpha_k}{2 \sin \alpha_k \cos 2l\alpha_k \cos (2l+2 ) \alpha_k} \\[.2cm] `\nonumber & = & \bar{x}_2 + (\bar{x}_2 - \bar{x}_1)\sum_{l=1}^{s-1} \frac{1-\theta_k}{1 - \frac{2}{1+\theta_k} \cos 2(2l +1)\alpha_k} .  
\end{eqnarray}
Recall that we aim to show $\eta_k < \bar{x}_3$ under the condition in \eqref{D22}, which by \eqref{J20} corresponds to
\begin{equation}
\label{J21}
\Delta_k :=\sum_{l=1}^{k-1} \frac{1-\theta_k}{1 + \frac{2}{1+\theta_k} \cos \frac{2l +1}{2k+1}\pi}\leq  \theta_k.
\end{equation}
By \eqref{1} and \eqref{M33}, we have
\begin{gather}
\label{J22}
1-\theta_k = 4 \sin^2\frac{\pi}{2(2k+1)}, \\[.2cm] \nonumber 1+\theta_k = - 2\cos \frac{2k\pi}{2k+1}= 2\cos \frac{\pi}{2k+1}.
\end{gather}
We use the latter in \eqref{J21} and obtain
\begin{gather}
\label{J24}
\Delta_k = \beta(u_k)\sum_{l=1}^{k-1}\gamma_l(u_k), \qquad u_k:= \frac{\pi}{2k+1}, \qquad k\geq 2  \\[.2cm] \nonumber\beta(u) = {2}\sin^2 {\frac{u}{2}} \cos u , \quad  \gamma_l(u) = \frac{1} {\cos lu \cos(l+1)u} , \quad u\in (0,u_k].
\end{gather}
Since $(l+1)u_k \leq  k\pi/(2k+1) < \pi/2$ for all $l\leq k-1$, it follows that
\begin{equation}
 \label{J24c}
 \gamma_l(u) >0, \quad {\rm for} \ \ {\rm all} \ \ l\leq k-1 \ \ {\rm and} \ \ u\leq u_k.
\end{equation}
Another convenient form is
\begin{equation}
\label{J25}
\Delta_{k} = \sum_{l=1}^{k-1} \frac{\beta(u_k)}{\sin\left(\frac{(2l-1)u_k}{2}\right) \sin\left(\frac{(2l+1)u_k}{2}\right) }. 
\end{equation}
Now we use in \eqref{J25} the following trigonometric identity
\begin{equation}
\label{I1}
\cot a - \cot b = \frac{\sin (b-a)}{\sin a \sin b},
\end{equation}
and obtain
\begin{eqnarray}
\label{I2}
\Delta_k & = & \frac{\beta(u_k)}{\sin u_k} \sum_{l=1}^{k-1} \left[ \cot\left(\frac{(2l-1)u_k}{2}\right) -  \cot\left(\frac{(2l+1)u_k}{2}\right) \right] \\[.2cm] \nonumber & =  &   \frac{\beta(u_k)}{\sin u_k} \left[ \cot\left(\frac{u_k}{2}\right) -  \cot\left(\frac{(2k-1)u_k}{2}\right) \right] \\[.2cm] \nonumber & =  &   \frac{\beta(u_k)\cos \left(\frac{u_k}{2}\right)}{\sin u_k \sin \left(\frac{u_k}{2}\right)} \left[1 - \tan u_k  \tan \left(\frac{u_k}{2}\right)  \right]\\[.2cm] \nonumber & =  & \cos u_k \left[1 - \tan u_k  \tan \left(\frac{u_k}{2}\right)  \right].
\end{eqnarray}
By \eqref{J24} and \eqref{I2}, and also by \eqref{1} and \eqref{M33}, we finally get
\begin{equation}
\label{I3}
\Delta_k  = \theta_k = \frac{\mathcal{M}_k-1}{\mathcal{M}_k + 1},
\end{equation}
which yields \eqref{J21}.
Now we calculate $t_k$. By \eqref{J2}, \eqref{J15}  and \eqref{J18}, it follows that
\begin{eqnarray}
\label{I4}
t_s & = & t_{s-1} + \frac{\delta_{s-1}(\alpha_k)}{v_{2,s-1}}(\bar{x}_2 - \bar{x}_1) =  
t_{s-1} - \frac{\sin 2\alpha_k \sin 4 \alpha_k (\bar{x}_2 - \bar{x}_1)}{\sin 4(s-1)\alpha_k \sin 4 s\alpha_k}\\ \nonumber & =: & t_{s-1} + (\bar{x}_2 - \bar{x}_1) \kappa_{s-1,k} , \qquad s \geq 2.
\end{eqnarray}
We iterate this recurrence, take $s=k$, use \eqref{I1}, and get
\begin{eqnarray}
\label{I5}
t_k & = & t_1 + (\bar{x}_2 - \bar{x}_1)\sum_{l=1}^{k-1} \kappa_{l,k} = t_1 - (\bar{x}_2 - \bar{x}_1)\sin 2 \alpha_k \sum_{l=1}^{k-1} \left[\cot 4l\alpha_k - \cot 4(l+1)\alpha_k\right]  \\[.2cm] \nonumber & = &  t_1 + (\bar{x}_2 - \bar{x}_1) \sin 2 \alpha_k    \left[\cot 4k\alpha_k - \cot 4\alpha_k\right].
\end{eqnarray}
Recall that $\alpha_k = k\pi /(2k+1)$, which means that $4k\alpha_k = 2k\pi - 2\alpha_k$. Then  take into account \eqref{M22} and \eqref{D1a}, and arrive at the following 
\begin{eqnarray}
\label{I6}
t_k & = & t_1 -(\bar{x}_2 - \bar{x}_1) \sin 2\alpha_k \left[\frac{\cos 2\alpha_k}{\sin 2\alpha_k}- \frac{2\cos^2 2\alpha_k-1}{2 \sin 2\alpha_k\cos 2\alpha_k}\right] \\[.2cm] \nonumber & = & t_1 + (\bar{x}_2 - \bar{x}_1)\left[ - 2 \cos 2\alpha_k + \frac{1}{ 2\cos 2\alpha_k}\right] \\[.2cm] \nonumber & = & \bar{x}_1 + \frac{\bar{x}_2-\bar{x}_1}{1+\theta_k} + (\bar{x}_2 - \bar{x}_1)\left[1+\theta_k - \frac{1}{1+\theta_k}\right]  \\[.2cm] \nonumber & = & \bar{x}_2 + \theta_k (\bar{x}_2 - \bar{x}_1).
\end{eqnarray}
By \eqref{J13}, then by \eqref{J8}, \eqref{J9}, and then by \eqref{J15} and some trigonometric identities, we obtain the following
\begin{eqnarray}
\label{I7}
\xi_k & = & \eta_k -\frac{v_{2,k}- v_{1,k}+ v_{2,k-1}}{(v_{0,k-1} - v_{1,k-1})(v_{2,k}-v_{1,k})} (\bar{x}_2 - \bar{x}_1)\\[.2cm] \nonumber & = & \eta_k - \frac{\sin^2 2\alpha_k \sin2k \alpha_k(\bar{x}_2 - \bar{x}_1)}{\sin^2 \alpha_k\sin 2(2k-1)\alpha_k\sin 4k \alpha_k} \\[.2cm] \nonumber & \times & \left[2\cos \alpha_k \cos 4 \alpha_k\sin 2k \alpha_k - \cos 2\alpha_k\sin (2k+1)\alpha_k\right] \\[.2cm] \nonumber & = & \eta_k - \frac{\cos \alpha_k \cos 2 \alpha_k}{\cos (2k-1)\alpha_k \cos 2k \alpha_k}(\bar{x}_2 - \bar{x}_1) \\[.2cm] \nonumber &  = & \eta_k -(\bar{x}_2 - \bar{x}_1) = \bar{x}_1 +\theta_k(\bar{x}_2 - \bar{x}_1), 
\end{eqnarray}
see \eqref{I3}. This completes the entire proof, see \eqref{D3w}.

\bibliography{references_cold_gas}
%\begin{thebibliography}{ll}
%\end{thebibliography}

\end{document}